\documentclass[a4paper,12pt]{article}
\usepackage{amssymb}
\usepackage{amsmath}
\usepackage{amstext}
\usepackage{amsfonts}
\usepackage{amsthm}
\usepackage{mathrsfs}
\usepackage{epsfig}
\usepackage{dcolumn}
\usepackage{rotating}
\usepackage{color}
\usepackage{comment}
\usepackage{hyperref}

\def\XXint#1#2#3{{\setbox0=\hbox{$#1{#2#3}{\int}$ }
\vcenter{\hbox{$#2#3$ }}\kern-.56\wd0}}

\newcommand*\xbar[1]{%
  \hbox{%
    \vbox{%
      \hrule height 0.5pt 
      \kern0.5ex
      \hbox{%
        \kern-0.1em
        \ensuremath{#1}%
        \kern-0.1em
      }%
    }%
  }%
}

\definecolor{rosso}{cmyk}{0,1,1,0.4}
\definecolor{rossos}{cmyk}{0,1,1,0.55}
\definecolor{rossoc}{cmyk}{0,1,1,0.2}
\definecolor{blu}{cmyk}{1,1,0,0.3}
\definecolor{blus}{cmyk}{1,1,0,0.6}
\definecolor{bluc}{cmyk}{1,1,0,0.1}
\definecolor{verde}{cmyk}{0.92,0,0.59,0.25}
\definecolor{verdec}{cmyk}{0.92,0,0.59,0.15}
\definecolor{verdes}{cmyk}{0.92,0,0.59,0.7}

\newcommand{\ba}{\begin{eqnarray}}
\newcommand{\ea}{\end{eqnarray}}
\newcommand{\be}{\begin{equation}}
\newcommand{\ee}{\end{equation}}
\newcommand{\bi}{\begin{itemize}}
\newcommand{\ei}{\end{itemize}}
\newcommand{\al}{\alpha}
\newcommand{\bt}{\beta}
\newcommand{\ga}{\gamma}

\newcommand{\la}{\lambda}

\newcommand{\sa}{\sigma}
\newcommand{\en}{\epsilon}

\newcommand{\Ga}{\Gamma}

\newcommand{\La}{\Lambda}

\newcommand{\cF}{{\cal F}}

\newcommand{\cO}{{\cal O}}

\newcommand{\w}{\widetilde}

\newcommand{\st}{\stackrel}

\newcommand{\ra}{\rightarrow}

\newcommand{\LF}{\left(}
\newcommand{\RF}{\right)}
\newcommand{\LT}{\left[}
\newcommand{\RT}{\right]}
\newcommand{\Ld}{\left.}
\newcommand{\Rd}{\right.}

\newcommand{\kb}{\bar{k}}
\newcommand{\pb}{\bar{p}}

\newcommand{\mt}{\mathtt}

\newcommand{\non}{\nonumber\\}

\usepackage[normalem]{ulem}

\begin{document}

\title{High-Energy Scatterings in Infinite-Derivative Field Theory and Ghost-Free Gravity}
\author{Spyridon Talaganis and Anupam Mazumdar\\\\
Consortium for Fundamental Physics, Lancaster University, LA1 4YB, UK}
\date{\today}

\maketitle

\begin{abstract}
In this paper, we will consider scattering diagrams in the context of infinite-derivative theories. First, we examine a finite-order, higher-derivative scalar field theory and find that we cannot eliminate the growth of scattering diagrams for large external momenta. Then, we employ an infinite-derivative scalar toy model and obtain that the external momentum dependence of scattering diagrams is convergent as the external momenta become very large. 
In order to eliminate the external momentum growth, one has to dress the bare vertices  of the scattering diagrams by considering renormalised propagator and vertex loop corrections to the bare vertices. Finally, we investigate scattering diagrams in the context of a scalar toy model which is inspired by a {\it ghost-free} and {\it singularity-free} infinite-derivative theory of gravity, where we conclude that infinite derivatives can eliminate the external momentum growth of scattering diagrams and make the scattering diagrams convergent in the ultraviolet.
\end{abstract}

\newpage

\setcounter{page}{1}

\tableofcontents

\section{Introduction}
\numberwithin{equation}{section}

Scattering diagrams play an important role in Quantum Field Theory (QFT). By studying scattering diagrams, one can obtain the scattering matrix element and, ultimately, the cross section. A cross section that blows up at high energies indicates an unphysical theory. 
Typically, in non-renormalisable theories, the cross section blows up at finite-order, see~\cite{Weinberg:1995mt}. For instance,  higher than $2$-derivative scalar field theories are one such example. Another example is indeed General Relativity (GR); also, in supergravity, see~\cite{Elvang:2013cua}, where high-energy scatterings of gravitons have been studied. Besides studying whether the amplitudes are finite or not, there are very interesting applications  in cosmology and in formation of mini black holes in trans-Plankian 
scatterings of plane waves~\cite{Amati0,Amati1,Amati2,Amati3,Amati4,Giddings0,Giddings1,Giddings:2001bu,Eardley:2002re}. In all these cases, the cross section of a scattering diagram, especially involving gravitons, blows up for large external momenta, {\it i.e.}, in the ultraviolet (UV). On the other hand, string theory has been conjectured to be UV-finite~\cite{Polchinski}; however, the problem here lies in higher-order corrections in string coupling $g_s$ and $\alpha'$, which would naturally induce corrections beyond Einstein-Hilbert action. Unfortunately, many of these corrections cannot be computed so easily in a time-dependent cosmological background. Nevertheless, there has been many studies in a fixed background in the context of string scatterings, see~\cite{Veneziano:1968yb,Gross:1987ar,Giddings:2007bw,Gross:1987kza}, see for details~\cite{Polchinski,Staessens:2010vi}.
Indeed, none of these analyses motivated from strings or supergravity can probe the region of space-time singularity; neither string theory nor supergravity in its current form can avoid forming a black hole or cosmological singularity. Besides string theory, there are other approaches of quantum gravity, such as in Loop Quantum Gravity (LQG)~\cite{Ashtekar,Nicolai:2005mc}, or in Causal Set approach~\cite{Henson:2006kf}, where it is possible to setup similar 
physical problems to study the behaviour at short distances and at small time scales, as well as high momentum scatterings.

One common thread in all these quantum and semiclassical approaches is the presence of non-locality, where the interactions happen in a finite region of spacetime. It has been conjectured by many that such non-local interactions may ameliorate the UV behaviour of scattering amplitudes, see~\cite{Giddings0,Giddings1,Giddings:2007bw,Gross:1987kza,Dona:2015tra,Veneziano1,Veneziano2,Veneziano3,Veneziano4,Veneziano5,Tseytlin:1995uq,Siegel:2003vt,Biswas:2004qu,Biswas:2014yia}, see also~Refs.~\cite{Biswas:2009nx,Biswas:2010xq,Biswas:2010yx,Biswas:2012ka} for finite temperature effects of non-local field theories. It is also expected that any such realistic theory of quantum gravity should be able to resolve short-distance and small-timescale singular behaviour present in Einstein's gravity, both in static and in time-dependent backgrounds. Indeed, close to the singularity or close to super-Planckian energies, one would naturally expect higher-derivative corrections to the Einstein-Hilbert action. Such higher-derivative corrections may as well open a door for non-local interactions in a very interesting way.

Typically, higher derivatives present a problem of {\it ghost}. For instance, it is well known that a quadratic curvature gravity is renormalisable, but would contain {\it ghosts} by virtue of having four derivatives in the equation of motion. The issue of ghost persists for any finite-order, higher than $2$-derivative theory for any spin. The issue of ghosts can be addressed in the context of an infinite-derivative~\footnote{Infinite derivatives are also present in (open) string field theory~\cite{Witten:1985cc} and in $p$-adic strings~\cite{Freund:1987kt}. The nonlocality of the invariant string field action was shown in~\cite{Gross:1986ia}. One would naturally expect them to be present from higher-order $\alpha'$ corrections.} theory of gravity, see Refs.~\cite{Tomboulis,Biswas:2005qr,Biswas:2011ar,Biswas:2013kla,Modesto}. The graviton propagator is definitely modified in this case as compared to the Einstein-Hilbert action. We should point out that infinite-derivative theories represent a novel approach of addressing some of the most important problems physics is facing. Among other things, the formulation of the initial value problem within the context of infinite-derivative theories remains a challenge;  in Ref.~\cite{Barnaby} it was shown that, in infinite-derivative theories, there are sometimes only two pieces of initial value data per pole under the assumption that temporal Fourier transforms exist. Numerically, one requires an ansatz to solve equations of motion containing infinite derivatives,  such as in the case of cosmology, see Ref.~\cite{Biswas:2005qr}. In this paper, we shall avoid these important issues by working perturbatively about a specific background in Euclidean momentum space.

In particular, in Ref.~\cite{Biswas:2011ar}, the authors constructed the most general {\it covariant} construction of quadratic-order gravity with infinite derivatives around Minkowski background. Similar construction is also possible around any {\it constant-curvature} backgrounds such as
in deSitter and (anti)-deSitter backgrounds~\cite{Biswas:2016etb}~\footnote{The quadratic curvature action is parity-invariant and torsion-free in both these cases~\cite{Biswas:2011ar,Biswas:2016etb}.}. In all these constructions~\cite{Biswas:2011ar,Biswas:2016etb}, it is possible to make the graviton propagator ghost-free, with no additional poles, other than the familiar $2$ massless degrees of freedom of Einstein-Hilbert action, by assuming that any modification which occurs as a result of infinite derivatives can be expressed by an 
{\it entire function}. An entire function as such does not introduce any pole in the complex plane. Furthermore, if the choice of 
an entire function is such that it falls off in the UV exponentially, while in the IR the function approaches unity in order to match the 
expectations of GR, then it can indeed soften the UV aspects of gravitational interactions. The fact that the propagator becomes exponentially suppressed in the UV, also leads to exponential enhancement in the vertex operator by virtue of derivative interactions. 
The interplay between the vertices and propagator give rise to this non-locality in gravity in the UV. Indeed, this non-locality is 
responsible for some nice properties, such as the resolution of cosmological and black hole type singularities~\footnote{In~\cite{Efimov}, one can see examples of non-local field theories which are not infinite-derivative ones; however, the approach cannot be helpful to address how to ameliorate the
singularity problems at short distances and small time scales.}.

For instance, it has been shown that for the above construction, it is possible to avoid cosmological singularity for a flat Universe 
Refs.~\cite{Biswas:2005qr,Biswas:2011ar,Biswas:2010zk,Biswas:2012bp,Conroy:2014dja}, which yields naturally a UV 
modification for Starobinsky inflation~\cite{Chialva:2014rla,Craps:2014wga}~\footnote{In GR, for a flat Universe it is extremely hard to avoid Big Bang singularity, the null congruence always
converge in a finite time~\cite{Borde:2001nh}, one requires softening of gravity in the UV in order to avoid cosmological singularity~\cite{Biswas:2005qr,Biswas:2011ar}.}. It is also possible to avoid
a black hole singularity in the linearised limit; the Newtonian potential is always finite in the UV in the limit $r\ra 0$, close to the source, see Ref.~\cite{Biswas:2011ar,Biswas:2013cha,Conroy:2015nva}. In Refs.~\cite{Frolov0,Frolov1,Frolov2}, authors have studied the time-dependent spherical 
collapse of matter for such non-local gravity~\cite{Biswas:2011ar}, and found that the singularity can be resolved at a linear regime.
Such time-dependent results are remarkable and clearly absent in Einstein Gravity and in finite-order higher-derivative 
modifications of gravity, such as in $4$th derivative gravity~\cite{Holdom:2002xy,Lu:2015psa}.

Furthermore, in Ref.~\cite{Talaganis:2014ida}, a {\it toy model} has been constructed inspired by an infinite-derivative extension of quadratic order gravity. Within this framework, quantum properties have been investigated, where UV divergences originating from Feynman diagrams have been studied {\it explicitly} up to $2$-loop order, and it was found that the Feynman diagrams become finite. A generic prescription was also provided  on how to make higher loops finite and, in fact, renormalisable~\cite{Talaganis:2014ida}.

Inspired by these recent developments, the aim of this paper is to study the high energy scatterings for ghost-free and infinite-derivative theories~\footnote{In principle, one can discuss breakdown of partial wave unitarity in Minkowski spacetime. A partial wave unitarity bound does not mean that beyond some energy scale unitarity is lost. It merely says that unitarity would be lost if perturbativity were assumed. In our case, we cannot define partial wave unitarity bound in Euclidean spacetime, as we shall see; instead we are keen to understand the scattering amplitudes which do not become arbitrarily large. This issue will become clear at later stages.} .We will study the $s$, $t$, $u$ channels of scattering diagrams  for a scalar field theory. In this respect we will be extending some of  earlier the computations of Refs.~\cite{Biswas:2014yia}~and~\cite{Talaganis:2014ida}. We will also study scattering diagrams within the scalar toy model of infinite-derivative quadratic curvature ghost-free and singularity-free gravity. In particular, we will show the following computations:

\begin{enumerate}

\item {\bf $2$-derivative massless scalar field, with higher derivative interactions:}
We will consider tree-level scattering diagrams, computed in Euclidean space. Then, we will look at the external 
momentum dependence of the scattering diagram if we insert a $1$-loop diagram in the middle. Next, we will replace 
the bare propagator in the tree-level diagram with the dressed one and see how the external momentum dependence 
of the diagram is modified. Finally, we will consider scattering diagrams with dressed vertices and propagators. In all cases, we will
find that the scattering  diagrams blow up in the UV limit. We will also compute the scattering diagrams by taking into account dressed propagators and dressed 
vertices, and the result would be the same.

\item {\bf Infinite-derivative Lagrangian and interactions:}
The results of the first computation motivate us to study a ghost-free, infinite-derivative Lagrangian with interaction terms containing infinite 
derivatives.  We will show that the scattering amplitude still blows up with and without dressed propagator. However, dressing the vertices by taking renormalised propagator and vertex loop corrections to the bare vertices eliminates the external momentum growth of the scattering amplitudes in the limit of the centre-of-mass energy going to infinity. 

\item {\bf Scalar toy model of infinite-derivative, ghost-free and singularity-free gravity:}
By taking the cue from our previous computations, we will then study a scalar toy model motivated from an infinite-derivative, ghost-free and singularity-free theory of gravity~\cite{Biswas:2005qr,Biswas:2011ar,Talaganis:2014ida}. We will show that a similar conclusion holds true for this class of action, where dressing the vertices by taking both propagator and vertex loop corrections to the bare vertices makes, at sufficiently high loop order, the external momentum dependence of any scattering diagram convergent in the UV.

\end{enumerate}



The paper is organised as follows: in section~\ref{sec:motiv}, we introduce a finite-order higher-derivative scalar field theory and examine the UV external momentum dependence of scattering diagrams. In section~\ref{sec:infdev}, we write down an infinite-derivative scalar field theory and study the external momentum dependence of scattering diagrams. In section~\ref{sec:infdevgrav}, we investigate external momentum dependence of scatterings of a scalar field theory analogue of infinite-derivative theory of gravity, and in section~\ref{sec:concl}, we conclude by summarising our results.


\section{Scatterings in scalar field theory with higher-derivative interactions}\label{sec:motiv}
\numberwithin{equation}{section}

\begin{figure}[t]
\centering
\includegraphics[width=.40\textwidth]{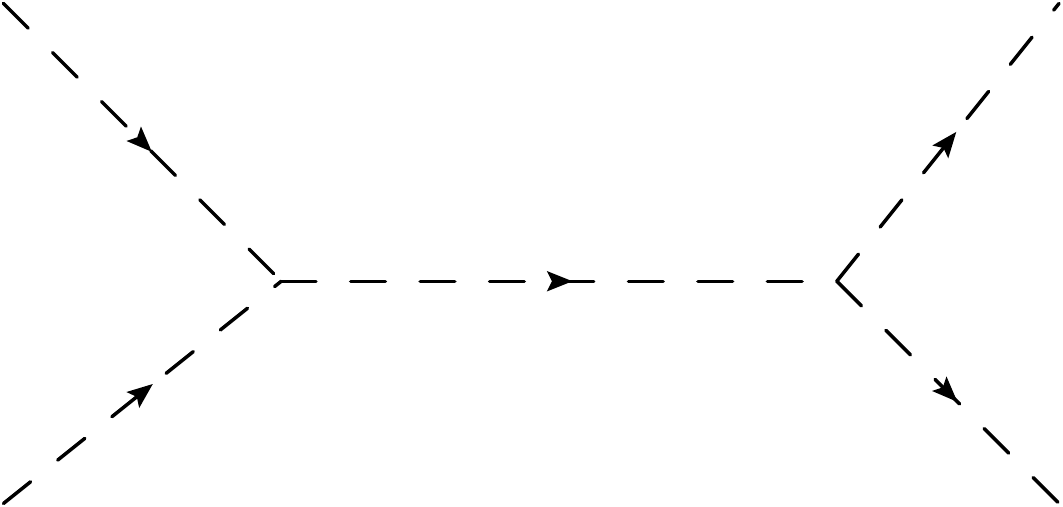}
\caption{\label{fig:tree} {\small The $s$-channel, tree-level scattering diagram $p_{1} p_{2} \ra p_{3} p_{4}$.}}
\end{figure}

\begin{figure}[t]
\centering
\includegraphics[width=.20\textwidth]{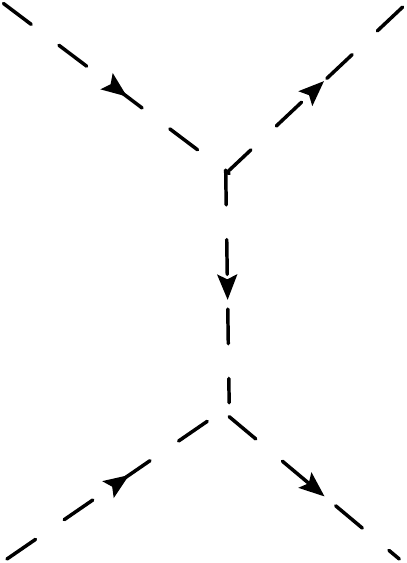}\hspace{3cm}
\includegraphics[width=.20\textwidth]{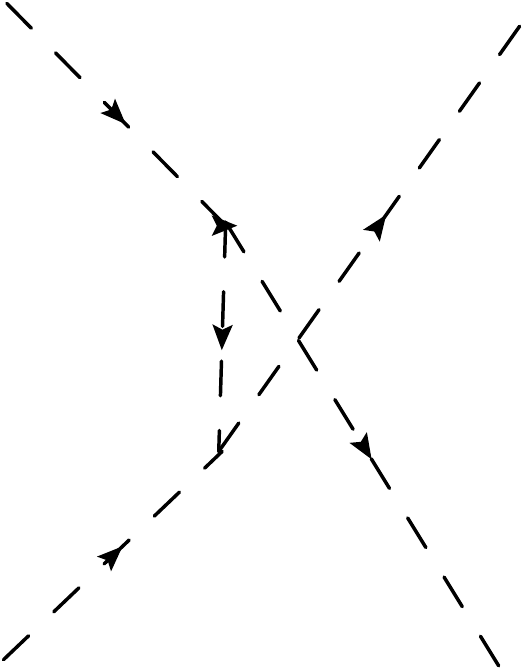}
\caption{\label{fig:st} {\small Left: The $t$-channel, tree-level scattering diagram $p_{1} p_{2} \ra p_{3} p_{4}$. Right: The $u$-channel, tree-level scattering diagram $p_{1} p_{2} \ra p_{3} p_{4}$.}}
\end{figure}

Let us now begin with a simple massless scalar field with a higher-derivative interaction term:
\be \label{eq:quad}
S=S_{\mt{free}}+S_{\mt{int}}\,,
\ee
where
\be
S_{\mt{free}}=\frac{1}{2}\int d^4 x \, \LF  \phi \Box \phi\RF
\ee
and
\be 
S_{\mt{int}}=\la \int d^4 x \, (\phi \Box \phi \Box \phi)\,,
\ee
where we treat $\lambda \ll {\cal O}(1)$, so that we are within the perturbative limit.
We will be working in an Euclidean space~\footnote{In Minkowski space (``mostly plus'' metric signature), $k^2 = - k_{0}^2 + \vec{k}^2$, where $\vec{k}^2 = k_{1}^{2}+k_{2}^{2}+k_{3}^{2}$. After analytic continuation, $k_{E}^2 = k_{4}^2 + \vec{k}^2$, where $k_{4} = -i k_{0}$. For brevity, we will suppress the subscript $E$ in the notations. For the rest of the paper we will continue our computations in Euclidean space.}, the propagator in the momentum space is then given by
\be \label{eq:propagator}
\Pi (k ^ 2)= \frac{- i}{k^2}\,,
\ee
while the vertex factor is given by:
\be \label{eq:vertex}
\la V(k_{1},k_{2},k_{3})=2i \la \left(k_{1}^{2}k_{2}^{2}+k_{2}^{2}k_{3}^{2}+k_{3}^{2}k_{1}^{2} \right) \,,
\ee
where
\be
k_{1}+k_{2}+k_{3}=0\,.
\ee
We can compute the tree-level amplitudes for the $s,~t,~u$ channels, see Fig.~\ref{fig:tree},
\be
i \mathcal{T}_{\mathrm{tree-level}}^{\mathrm{s-channel}}=-\frac{25}{4}\la^{2} s^{4}\LF \frac{i}{s} \RF\,,
\ee
where $s=-(p_{1}+p_{2})^2$. Similarly, see Fig.~\ref{fig:st} (left),
 \be
i \mathcal{T}_{\mathrm{tree-level}}^{\mathrm{t-channel}}=-4\la^{2}s^{2}\left(t+\frac{s}{4}\right)^{2}\LF \frac{i}{t} \RF
\ee
and, see Fig.~\ref{fig:st} (right),
\be
i \mathcal{T}_{\mathrm{tree-level}}^{\mathrm{u-channel}}=-4\la^{2}s^{2}\left(u+\frac{s}{4}\right)^{2}\LF \frac{i}{u} \RF\,,
\ee
where $t=-(p_{1}-p_{3})^2$ and $u=-(p_{1}-p_{4})^2$. Hence, the total amplitude is given by:
\be\label{eq:210}
\mathcal{T}_{\mathrm{tree-level}}=-4\la^{2}s^{2}\LF \frac{\left(\frac{5s}{4}\right)^2}{s}+\frac{\left(t+\frac{s}{4}\right)^2}{t}+\frac{\left(u+\frac{s}{4}\right)^2}{u} \RF\,.
\ee
Since the scattering matrix element $\mathcal{T}_{\mathrm{tree-level}}$ in Eq.~\eqref{eq:210} blows up as $s \ra -\infty$, the total cross section $\sa_{\mathrm{tree-level}}$ in the centre-of-mass (CM) frame (see Eqs.~\eqref{eq:sigma}~\&~\eqref{eq:a4} in appendix~\ref{sec:defconv} for the definition of $\sa$) also blows up as $s \ra - \infty$.

\subsection{Dressing the propagator}

\begin{figure}[t]
\centering
\includegraphics[width=.80\textwidth]{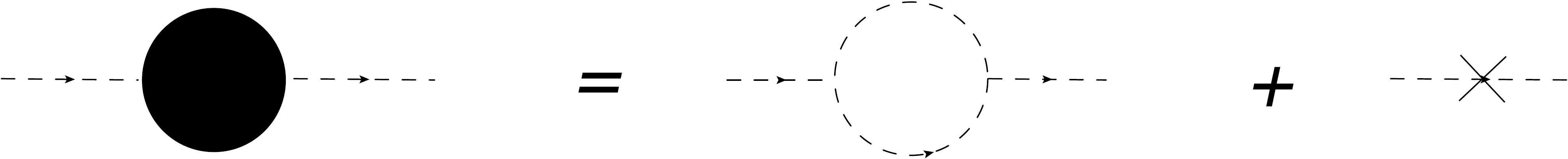}
\caption{\label{fig:okopoko} {\small The $1$-loop, $2$-point contribution of $1$PI diagrams. The cross denotes a counter-term vertex.}}
\end{figure}

\begin{figure}[t]
\centering
\includegraphics[width=.50\textwidth]{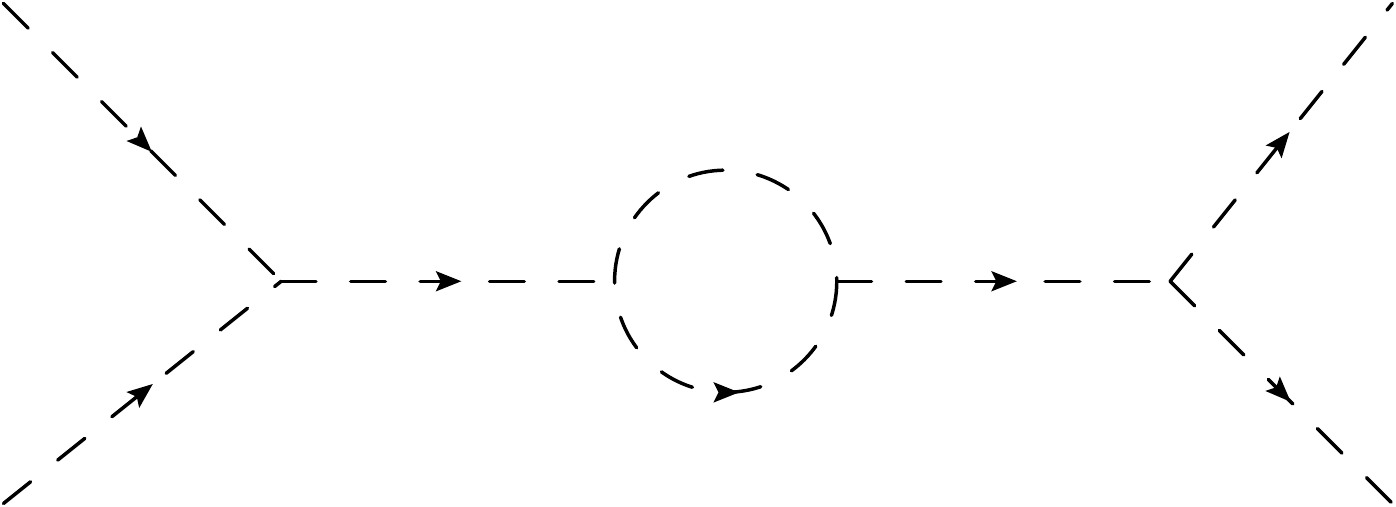}
\caption{\label{fig:kuk} {\small The $s$-channel, $1$-loop scattering diagram $p_{1} p_{2} \ra p_{3} p_{4}$.}}
\end{figure}

\begin{figure}[t]
\centering
\includegraphics[width=1.00\textwidth]{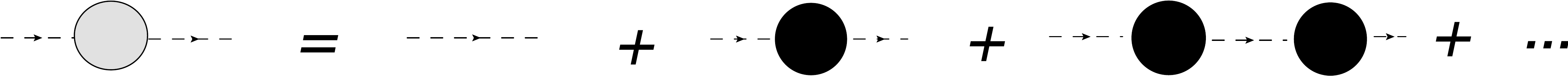}\vspace{1cm}
\includegraphics[width=.40\textwidth]{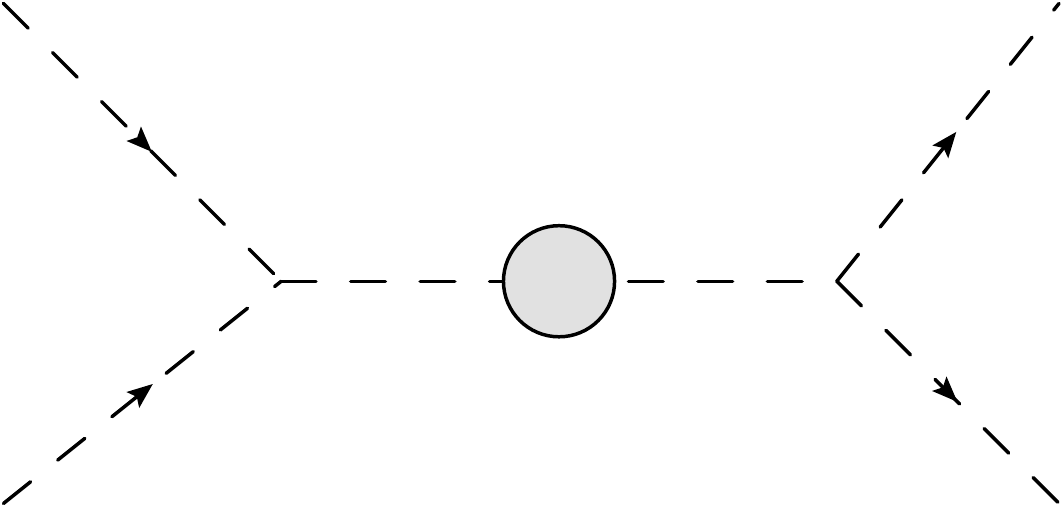}
\caption{\label{fig:dressed} {\small Top: The dressed propagator as the sum of an infinite geometric series. The dressed propagator is denoted by the shaded blob. Bottom: The $s$-channel, scattering diagram $p_{1} p_{2} \ra p_{3} p_{4}$ in which the bare propagator is replaced by the dressed propagator. The shaded blob indicates a dressed propagator.}}
\end{figure}

Since the tree-level amplitude blows up, we should now study the $1$-loop, $2$-point function in the propagator for the above interaction, see Eq.~\eqref{eq:quad}. We can compute the $1$-loop, $2$-point function with arbitrary external momentum, $p$. Therefore, regarding the $1$-loop, $2$-point function with external momenta $p$, $-p$ and symmetrical routing of momenta, see Fig.~\ref{fig:okopoko}, we have
\begin{align}\label{eq:lambda}
\Ga_{2,1}(p^2)& = \frac{i \la^2}{2} \int^{\La} \frac{d^4k}{(2\pi)^4} \, \frac{4\LT p^{2}(\frac{p}{2}-k)^{2}+p^{2}(\frac{p}{2}+k)^{2}+(\frac{p}{2}+k)^{2}(\frac{p}{2}-k)^{2}\RT^2}{(\frac{p}{2}-k)^{2}(\frac{p}{2}+k)^{2}}  \non
& = \frac{i\la^{2}}{2} \int_{0}^{\La}d\mathbf{k} \int_{-1}^{1}dx \, \frac{4\pi\mathbf{k}^{3}\sqrt{1-x^{2}}}{(2\pi)^{4}} \frac{4\LT p^{2}(\frac{p}{2}-k)^{2}+p^{2}(\frac{p}{2}+k)^{2}+(\frac{p}{2}+k)^{2}(\frac{p}{2}-k)^{2}\RT^2}{(\frac{p}{2}-k)^{2}(\frac{p}{2}+k)^{2}}\non
&= i \la^{2} \left(-\frac{p^8}{48 \pi ^2}+\frac{\Lambda ^2 p^6}{8 \pi ^2}+\frac{81 \Lambda ^4 p^4}{256 \pi ^2}+\frac{17 \Lambda ^6 p^2}{96 \pi ^2}+\frac{\Lambda ^8}{32 \pi ^2}\right)\,,
\end{align}
where $k$ is the internal loop momentum, $x$ is the cosine of the angle between $p$ and $k$ ($p \cdot k = \mathbf{p}\,\mathbf{k}\,x$, where $\mathbf{p}$ and $\mathbf{k}$ are the norms of $p$ and $k$ in Euclidean space) and $\Lambda$ is a hard cutoff. The counter-term, which is needed to cancel the divergences denoted by powers of $\Lambda$ in Eq.~\eqref{eq:lambda}, and which should be added to the action in Eq.~\eqref{eq:quad}, is given by
\be
S_{\mt{ct}}=\frac{\la^{2}\La^{2}}{16\pi^{2}}\int d^{4}x \, \LF \phi \Box^{3}\phi-\frac{81\La^{2}}{32}\phi \Box^{2} \phi+\frac{17\La^{4}}{12}\phi \Box \phi-\frac{\La^{6}}{4}\phi^{2} \RF \,,
\ee 
which yields
\be
\Ga_{2,1,\mt{ct}}(p^{2})=-\frac{i\la^{2}\La^{2}}{8\pi^{2}}\left(p^{6}+\frac{81 \Lambda ^2 p^4}{32}+\frac{17 \Lambda ^4 p^2}{12}+\frac{\Lambda ^6}{4} \right)\,.
\ee 
Thus, the renormalised $1$-loop, $2$-point function is 
\be\label{eq:ren}
\Ga_{2,1\mt{r}}(p^{2})=\Ga_{2,1}(p^{2})+\Ga_{2,1,\mt{ct}}(p^{2})=-\frac{i\la^{2}p^{8}}{48\pi^{2}}\,.
\ee
We observe that the maximum power of $p$ appearing in the renormalised $1$-loop, $2$-point function with arbitrary external momenta, Eq.~\eqref{eq:ren}, is $p^8$. Hence, in the UV, {\it i.e.}, in the limit $s\ra -\infty$, $\Ga_{2,1\mt{r}}(-s) \propto (p_{1}+p_{2})^{8} = s^{4}$, where $\Ga_{2,1\mt{r}}$ is the renormalised $1$-loop, $2$-point function. Since
\begin{align}
i\mathcal{T}_{\mathrm{1-loop}} &=  \la^{2} (p_{1},p_{2},-p_{1}-p_{2}) V(-p_{3},-p_{4},p_{1}+p_{2}) \LF \frac{i}{s} \RF ^ {2} \Ga_{2,1\mt{r}}(-s)  \non
&+ \la^{2} V(p_{1},-p_{3},p_{3}-p_{1}) V(p_{2},-p_{4},p_{1}-p_{3}) \LF \frac{i}{t } \RF ^ {2} \Ga_{2,1\mt{r}}(-t)  \non
&+ \la^{2} V(p_{1},-p_{4},p_{4}-p_{1}) V(p_{2},-p_{3},p_{1}-p_{4}) \LF \frac{i}{u } \RF ^ {2}\Ga_{2,1\mt{r}}(-u) \,,
\end{align} 
the $s$-channel of $\mathcal{T}_{\mathrm{1-loop}}$ goes as  $s^{2}s^{2}s^{-2}s^{4}=s^{6}$ when $s \ra -\infty$, see Fig.~\ref{fig:kuk} (the two bare propagators go as $1/s$ each while the two bare vertices go as $s^{2}$ each). Hence, as $s \ra - \infty$, $\mathcal{T}_{\mathrm{1-loop}}^{\mathrm{s-channel}}$ diverges. $\mathcal{T}_{\mathrm{1-loop}}^{\mathrm{t-channel}}$ and $\mathcal{T}_{\mathrm{1-loop}}^{\mathrm{u-channel}}$ also diverge except for $\theta=0$ and $\theta=\pi$, respectively. 

Now what if we had an infinite series of loops in the scattering diagrams, see Fig.~\ref{fig:dressed} (top), that is, if we had replaced the bare propagator with the {\it dressed propagator}?
As we shall see below, the external momentum dependence of the $1$-loop, $2$-point function shall actually determine the UV behaviour of the dressed propagator.

The dressed propagator, see Fig.~\ref{fig:dressed} (top), represents the geometric series of all the graphs with $1$-loop, $2$-point insertions, analytically continued to the entire complex $p^2$-plane.  Mathematically, the dressed propagator, $\w{\Pi}(p^2)$, is given by~\cite{Talaganis:2014ida}
\be
\w{\Pi}(p^2)= \frac{\Pi(p^2)}{1-\Pi(p^2)\Ga_{2,1\mt{r}}(p^2)}\,.
\ee
Hence, for our example, we have
\begin{align}\label{eq:den}
\w{\Pi}(p^2)&= \frac{-\frac{i}{p^2}}{1-\left( -\frac{i}{p^2} \right) \LF -\frac{i\la^{2}p^{8}}{48\pi^{2}} \RF} \non
&=\frac{-i}{p ^ {2 } +\frac{\la^{2} p^8}{48 \pi ^2}}\,.
\end{align}
For large $p$, $p^8$ dominates $p^2$ in the denominator of Eq.~\eqref{eq:den}, and we have
\be
\w{\Pi}(p^2) \approx - \frac{48 \pi^{2}i}{\la^{2} p^{8}}\,.
\ee

Since
\begin{align}
i\mathcal{T}_{\mathrm{dressed}} &= \la^{2} V(p_{1},p_{2},-p_{1}-p_{2}) V(-p_{3},-p_{4},p_{1}+p_{2}) \w{\Pi}(-s) \non
&+ \la^{2}  V(p_{1},-p_{3},p_{3}-p_{1}) V(p_{2},-p_{4},p_{1}-p_{3}) \w{\Pi}(-t)  \non
&+ \la^{2}  V(p_{1},-p_{4},p_{4}-p_{1}) V(p_{2},-p_{3},p_{1}-p_{4}) \w{\Pi}(-u) \,,
\end{align}
then, if we replace the bare propagator with the dressed propagator in the tree-level scattering diagrams, see Fig.~\ref{fig:dressed} (bottom), we will have
\begin{align}
\mathcal{T}_{\mathrm{dressed}}^{\mathrm{s-channel}}&=-\frac{25}{4}\la^2 \frac{s^3}{1-\frac{\la^2 s^3}{48\pi^2}}\,,\\
\mathcal{T}_{\mathrm{dressed}}^{\mathrm{t-channel}}&=-4\la^{2} \LF \frac{3s}{4}-\frac{s}{2}\cos \theta \RF^{2} \frac{2s}{(1-\cos \theta)\LT 1-\frac{\la^2 s^3(1-\cos \theta)^3}{384\pi^2}\RT}  \,,\\
\mathcal{T}_{\mathrm{dressed}}^{\mathrm{u-channel}}&=-4\la^{2} \LF \frac{3s}{4}+\frac{s}{2}\cos \theta \RF^{2} \frac{2s}{(1+\cos \theta)\LT 1-\frac{\la^2 s^3(1+\cos \theta)^3}{384\pi^2}\RT}\,.
\end{align}
Hence, we can make the following observations:
\begin{itemize}

\item{$\mathcal{T}_{\mathrm{dressed}}^{\mathrm{s-channel}}$ does not blow up as $s \ra - \infty$.}

\item{$\mathcal{T}_{\mathrm{dressed}}^{\mathrm{t-channel}}$ blows up as $s \ra -\infty$ when $\cos(\theta)=1 \Rightarrow \theta=0$.}

\item{Similarly, $\mathcal{T}_{\mathrm{dressed}}^{\mathrm{u-channel}}$ blows up as $s \ra -\infty$ when $\cos(\theta)=-1 \Rightarrow \theta=\pi$.}

\end{itemize}

Since we have that $\mathcal{T}_{\mathrm{dressed}}=\mathcal{T}_{\mathrm{dressed}}^{\mathrm{s-channel}}+\mathcal{T}_{\mathrm{dressed}}^{\mathrm{t-channel}}+\mathcal{T}_{\mathrm{dressed}}^{\mathrm{u-channel}}$, one can verify that the total cross section $\sigma_{\mathrm{dressed}}$ corresponding to $\mathcal{T}_{\mathrm{dressed}}$ blows up as $s \ra - \infty$. The summary is that the dressed propagator is not sufficient to prevent the 
scattering diagram from blowing up as $s \ra -\infty$ since the polynomial suppression coming from the dressed propagator cannot overcome the polynomial enhancement originating from the two bare vertices in Fig.~\ref{fig:dressed} (bottom). In subsection~\ref{sec:dv1}, we shall dress the vertices to see whether we can eliminate the external momentum divergences of the scattering diagrams.


\subsubsection{$1$-loop, $3$-point diagram with bare vertices and bare propagators}\label{tri}

\begin{figure}[t]
\centering
\includegraphics[width=.40\textwidth]{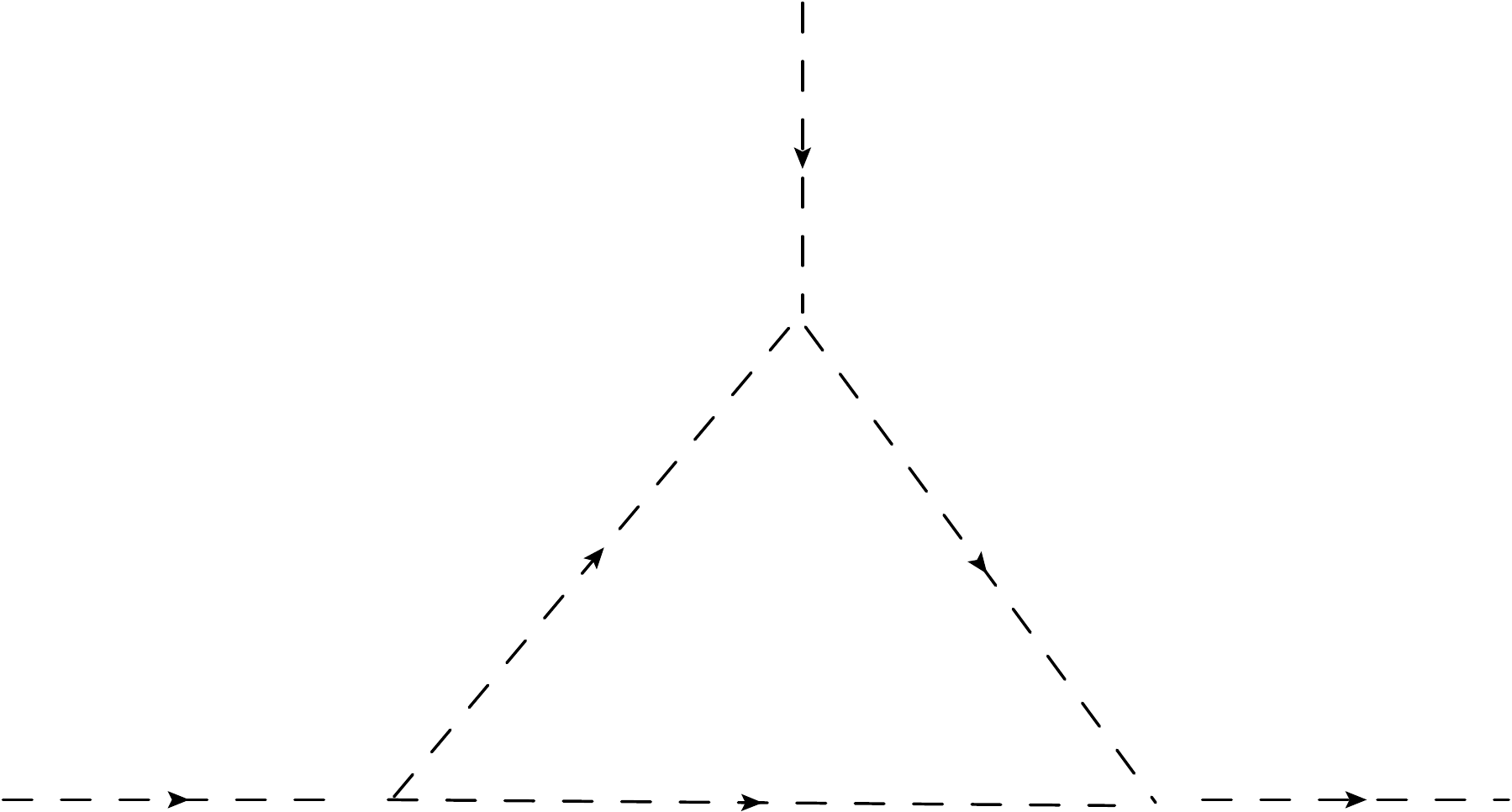}
\caption{\label{fig:set2} {\small $1$-loop, $3$-point diagram with bare vertices and bare internal propagators and symmetrical routing of momenta. The external momenta are $p_1$, $p_2$, $p_3$ and the internal (that is, inside the loop) momenta are $k+\frac{p_{1}}{3}-\frac{p_2}{3}$, $k+\frac{p_{2}}{3}-\frac{p_3}{3}$, $k+\frac{p_{3}}{3}-\frac{p_1}{3}$.}}
\end{figure}

As a prelude to subsection~\ref{sec:dv1}, suppose we consider a $1$-loop, $3$-point diagram, see Fig.~\ref{fig:set2}, with external momenta $p_{1}$, $p_{2}$ and $p_{3}$ (we assume that the propagators and the vertices are bare), and symmetrical routing of momenta. Then the propagators in the $1$-loop triangle are given by Eq.~\eqref{eq:propagator}:
\be \label{eq:218}
-i\LF k + \frac{p_{1}}{3} - \frac{p _ {2}}{3} \RF^{-2}\,, -i \LF k + \frac{p_{2}}{3} - \frac{p _ {3}}{3} \RF^{-2}\,, -i \LF k + \frac{p_{3}}{3} - \frac{p _ {1}}{3} \RF^{-2}\,, 
\ee
and the vertex factors are given by Eq.~\eqref{eq:vertex}: 
\ba
&& 2 i \la \LF p_{2}^{2}\LF k + \frac{p_{1}}{3} - \frac{p _ {2}}{3} \RF^{2}+p_{2}^2\LF k + \frac{p_{2}}{3} - \frac{p _ {3}}{3} \RF^{2}+\LF k + \frac{p_{1}}{3} - \frac{p _ {2}}{3} \RF^{2}\LF k + \frac{p_{2}}{3} - \frac{p _ {3}}{3} \RF^{2} \RF \,, \non 
&& 2 i \la \LF p_{3}^{2}\LF k + \frac{p_{2}}{3} - \frac{p _ {3}}{3} \RF^{2}+p_{3}^2\LF k + \frac{p_{3}}{3} - \frac{p _ {1}}{3} \RF^{2}+\LF k + \frac{p_{2}}{3} - \frac{p _ {3}}{3} \RF^{2}\LF k + \frac{p_{3}}{3} - \frac{p _ {1}}{3} \RF^{2} \RF  \,,\non 
&& 2 i \la \LF p_{1}^{2}\LF k + \frac{p_{3}}{3} - \frac{p _ {1}}{3} \RF^{2}+p_{1}^2\LF k + \frac{p_{1}}{3} - \frac{p _ {2}}{3} \RF^{2}+\LF k + \frac{p_{3}}{3} - \frac{p _ {1}}{3} \RF^{2}\LF k + \frac{p_{1}}{3} - \frac{p _ {2}}{3} \RF^{2} \RF \,.~~~~~~~~~
\ea 
Hence, the $1$-loop, $3$-point diagram, $\Ga_{3,1}(p^2)$, will be given by
\begin{align}\label{eq:31}
\Ga_{3,1}(p^2)&=i\la^{3}\int ^{\La} \frac{d^4 k}{(2 \pi)^4} \, \LT \frac{8\LF p_{2}^{2}\LF k + \frac{p_{1}}{3} - \frac{p _ {2}}{3} \RF^{2}+p_{2}^2\LF k + \frac{p_{2}}{3} - \frac{p _ {3}}{3} \RF^{2}+\LF k + \frac{p_{1}}{3} - \frac{p _ {2}}{3} \RF^{2}\LF k + \frac{p_{2}}{3} - \frac{p _ {3}}{3} \RF^{2} \RF}{\LF k + \frac{p_{1}}{3} - \frac{p _ {2}}{3} \RF^{2}\LF k + \frac{p_{2}}{3} - \frac{p _ {3}}{3} \RF^{2}\LF k + \frac{p_{3}}{3} - \frac{p _ {1}}{3} \RF^{2}}\Rd \non
& \times \LF p_{3}^{2}\LF k + \frac{p_{2}}{3} - \frac{p _ {3}}{3} \RF^{2}+p_{3}^2\LF k + \frac{p_{3}}{3} - \frac{p _ {1}}{3} \RF^{2}+\LF k + \frac{p_{2}}{3} - \frac{p _ {3}}{3} \RF^{2}\LF k + \frac{p_{3}}{3} - \frac{p _ {1}}{3} \RF^{2} \RF \non
& \times \Ld \LF p_{1}^{2}\LF k + \frac{p_{3}}{3} - \frac{p _ {1}}{3} \RF^{2}+p_{1}^2\LF k + \frac{p_{1}}{3} - \frac{p _ {2}}{3} \RF^{2}+\LF k + \frac{p_{3}}{3} - \frac{p _ {1}}{3} \RF^{2}\LF k + \frac{p_{1}}{3} - \frac{p _ {2}}{3} \RF^{2} \RF \vphantom{\frac{8\LF p_{2}^{2}\LF k + \frac{p_{1}}{3} - \frac{p _ {2}}{3} \RF^{2}+p_{2}^2\LF k + \frac{p_{2}}{3} - \frac{p _ {3}}{3} \RF^{2}+\LF k + \frac{p_{1}}{3} - \frac{p _ {2}}{3} \RF^{2}\LF k + \frac{p_{2}}{3} - \frac{p _ {3}}{3} \RF^{2} \RF}{\LF k + \frac{p_{1}}{3} - \frac{p _ {2}}{3} \RF^{2}\LF k + \frac{p_{2}}{3} - \frac{p _ {3}}{3} \RF^{2}\LF k + \frac{p_{3}}{3} - \frac{p _ {1}}{3} \RF^{2}}} \RT\,.
\end{align} 
After integration with respect to the internal loop momentum $k$ and renormalisation of the loop integral divergences, {\it i.e.}, the terms involving powers of $\La$, we are left with a polynomial function of the three external momenta $p_{1}$, $p_{2}$, $p_{3}$. We will require these computations in the following subsection.


\subsection{Dressing the vertices by making vertex loop corrections to the bare vertices}\label{sec:dv1}

\begin{figure}[t]
\centering
\includegraphics[width=.40\textwidth]{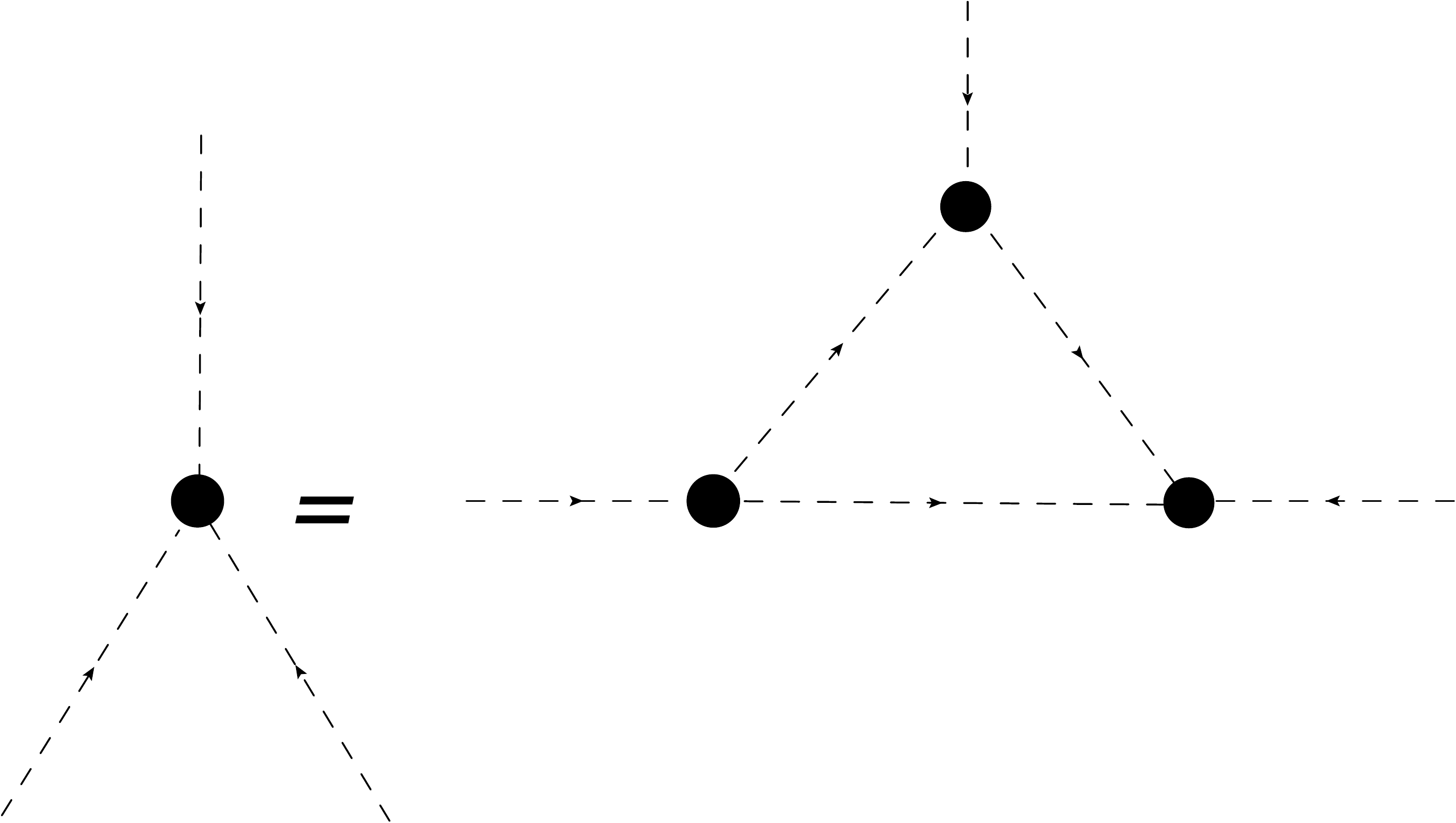}
\caption{\label{fig:set22} {\small $3$-point diagram constructed out of lower-loop $2$-point \& $3$-point diagrams. The dark blobs indicate renormalised vertex corrections and the dashed lines inside the triangle denote bare internal propagators.  The loop order of the dark blob on the left is $n$ while the loop order of the dark blobs on the right is $n-1$. The external momenta are $p_1$, $p_2$, $p_3$ and the internal (that is, inside the loop) momenta are $k+\frac{p_{1}}{3}-\frac{p_2}{3}$, $k+\frac{p_{2}}{3}-\frac{p_3}{3}$, $k+\frac{p_{3}}{3}-\frac{p_1}{3}$.}}
\end{figure}

Based on the results of subsection~\ref{tri}, suppose we want to dress the vertices by making renormalised vertex loop corrections to the bare vertices at the left- and right-ends of the scattering diagrams, see Fig.~\ref{fig:set22}. As we saw in Eq.~\eqref{eq:31}, both the bare propagators and the bare vertices can be written in terms of powers of momenta. After integration with respect to the internal loop momentum $k$, we obtain a polynomial expression involving powers of the external momenta $p_1$, $p_2$, $p_3$. As the loop-order increases, the $3$-point function can still be written as a polynomial function of the external momenta; this happens because, as previously, the (bare) propagators are polynomials in momenta while the (dressed) vertices are also polynomials in momenta. Therefore, we expect the external momentum dependence of the $3$-point function, see Fig~\ref{fig:set22}, in the UV limit, {\it i.e.}, as $p_{i}\ra \infty$, where $i=1,2,3$, in terms of the three external momenta, $p_1$, $p_2$, $p_3$, to follow as:
\be
\label{eq:popopo}
\Ga_3\st{UV}{\longrightarrow}\sum_{\al,\bt,\ga} p_{1}^{2\al}p_{2}^{2\bt}p_{3}^{2\ga} \,,
\ee
with  the convention
\be
\al\geq\bt\geq\ga \,.
\ee
The reason we expect the external momentum dependence of $3$-point function to be given by Eq.~\eqref{eq:popopo} is that, once all the (lower-) loop subdiagrams have been integrated out, what remains are polynomial expressions in terms of the corresponding external momenta. Some of these external momenta can then become the internal loop momentum in a subsequent higher-loop diagram.

First, let us consider how one can get the largest sum of all the exponents, {\it i.e.}, $\al+\bt+\ga$. Although all the arguments below can be conducted for three different sets of exponents in the three $3$-point vertices making up the $1$-loop triangle, see Fig.~\ref{fig:set22}, for simplicity, here we will look at what happens when all the three vertices have the same exponents.

Clearly, the best way to obtain the largest exponents for the external momenta is to have the $\al$ exponent correspond to the external momenta. Assuming a symmetric distribution of $(\bt, \ga)$ among the internal loops and considering the $n$-loop, $3$-point diagram with symmetrical routing of momenta,  see Fig.~\ref{fig:set22}, the propagators in the $1$-loop triangle are given by Eq.~\eqref{eq:218} and the vertex factors are~\footnote{The superscripts in $\al^{n-1}$, $\bt^{n-1}$, $\ga^{n-1}$ denote the loop-order; clearly, they are not powers.} 
\ba
&& i p _ {1} ^ {2\al^{n-1}} \LF k + \frac{p _ {3}}{3} - \frac{p _ {1}}{3} \RF^{2\bt^{n-1}} \LF k+\frac{p _ {1}}{3} - \frac{p _ {2}}{3} \RF ^{2\ga^{n-1}}\,, \non 
&& i p _ {2} ^ {2\al^{n-1}} \LF k + \frac{p _ {1}}{3} - \frac{p _ {2}}{3} \RF^{2\bt^{n-1}} \LF k + \frac{p _ {2}}{3} - \frac{p _ {3}}{3} \RF ^{2\ga^{n-1}}\,,\non && i p _ {3} ^ {2\al^{n-1}} \LF k + \frac{p _ {2}}{3} - \frac{p _ {3}}{3} \RF^{2\bt^{n-1}} \LF k + \frac{p _ {3}}{3} - \frac{p _ {1}}{3} \RF ^{2\ga^{n-1}}\,.
\ea
 
Conservation of momenta then yields, in the UV, {\it i.e.}, as $p_{i} \ra \infty$, where $i=1,2,3$,
\begin{align}\label{eq:www}
\Ga_{3,n}&{\longrightarrow}\int \frac{\mathrm{d} ^ 4 k}{(2 \pi) ^ 4} \, \LT \frac{p_{1}^{2\al^{n-1}}p_{2}^{2\al^{n-1}}p_{3}^{2\al^{n-1}}}{\LF k + \frac{p_{1}}{3} - \frac{p _ {2}}{3} \RF^{2}\LF k + \frac{p_{2}}{3} - \frac{p _ {3}}{3} \RF^{2}\LF k + \frac{p_{3}}{3} - \frac{p _ {1}}{3} \RF^{2}}\Rd \non
& \times \Ld \LF k + \frac{p_{1}}{3} - \frac{p _ {2}}{3} \RF^{2(\bt^{n-1}+\ga^{n-1})} \LF k + \frac{p_{2}}{3} - \frac{p _ {3}}{3} \RF^{2(\bt^{n-1}+\ga^{n-1})}\LF k + \frac{p_{3}}{3} - \frac{p _ {1}}{3} \RF^{2(\bt^{n-1}+\ga^{n-1})}\vphantom{\frac{p_{1}^{2\al^{n-1}}p_{2}^{2\al^{n-1}}p_{3}^{2\al^{n-1}}}{\LF k + \frac{p_{1}}{3} - \frac{p _ {2}}{3} \RF^{2}\LF k + \frac{p_{2}}{3} - \frac{p _ {3}}{3} \RF^{2}\LF k + \frac{p_{3}}{3} - \frac{p _ {1}}{3} \RF^{2}}}\RT\,,
\end{align}
where $p_{1}$, $p_{2}$, $p_{3}$ are the external momenta for the $1$-loop triangle and the superscript in the $\al,\bt,\ga$ indicates that these are coefficients that one obtains from contributions up to $n-1$  loop level.
Now, let us proceed to obtain the $n$-th loop coefficients. We can read from Eq.~(\ref{eq:www}):
\be \label{eq:abc}
\al^n=\bt^n=\ga^n=\al^{n-1}+2(\bt^{n-1}+\ga^{n-1})\,.
\ee
For $3$-point bare vertices, we have now $\al^{0}=\bt^{0}=1$ and $\ga^{0}=0$. As $n$ increases, $\al^n$, $\bt^n$ and $\ga^n$ increase; this means that, as the number of loops increases, the external momentum dependences of the dressed vertices become larger and larger as the external momenta become larger.

If we now dress the vertices by making renormalised vertex loop corrections to the bare vertices at the left- and right-ends of the tree-level scattering diagrams, we will
have, see Fig.~\ref{fig:achacha} (for $n \geq 1$, $\al^{n}=\bt^{n}=\ga^{n}$),
\begin{align}
\mathcal{T}_{\mathrm{vertex~corrections}}^{\mathrm{s-channel}}& \sim s^{2\al^{n}} \LF \frac{s}{2} \RF ^{4\al^{n}} \frac{1}{s} \,,\\
\mathcal{T}_{\mathrm{vertex~corrections}}^{\mathrm{t-channel}}& \sim t^{2\al^{n}} \LF \frac{s}{2} \RF ^{4\al^{n}} \frac{1}{t}= \LT \frac{s}{2}(1-\cos \theta) \RT ^{2\al^{n}-1} \LF \frac{s}{2} \RF ^{4\al^{n}} \,,\\
\mathcal{T}_{\mathrm{vertex~corrections}}^{\mathrm{u-channel}}& \sim u^{2\al^{n}} \LF \frac{s}{2} \RF ^{4\al^{n}} \frac{1}{u}=\LT \frac{s}{2}(1+\cos \theta) \RT ^{2\al^{n}-1} \LF \frac{s}{2} \RF ^{4\al^{n}} \,.
\end{align} 
Since $\al^{0}=\bt^{0}=1$ and $\ga^{0}=0$, using Eq.~\eqref{eq:abc}, we can see that $\al^{1}=3$; therefore, $\al^{n} \geq 3$ for $n \geq 1$. Hence, we can make the following observations from the above expressions:

\begin{itemize}

\item{$\mathcal{T}_{\mathrm{vertex~corrections}}^{\mathrm{s-channel}}$ blows up as $s \ra - \infty$.}

\item{$\mathcal{T}_{\mathrm{vertex~corrections}}^{\mathrm{t-channel}}$ blows up as $s \ra -\infty$ except when $\cos(\theta)=1 \Rightarrow \theta=0$.}

\item{Similarly, $\mathcal{T}_{\mathrm{vertex~corrections}}^{\mathrm{u-channel}}$ blows up as $s \ra -\infty$ except when $\cos(\theta)=-1 \Rightarrow \theta=\pi$.}

\end{itemize}
Thus, one can check that the cross section $\sa_{\mathrm{dressed~vertices}}$ corresponding to $\mathcal{T}_{\mathrm{vertex~corrections}}=\mathcal{T}_{\mathrm{vertex~corrections}}^{\mathrm{s-channel}}+\mathcal{T}_{\mathrm{vertex~corrections}}^{\mathrm{t-channel}}+\mathcal{T}_{\mathrm{vertex~corrections}}^{\mathrm{u-channel}}$ blows up as $s \ra - \infty$. 

We see that dressing the vertices by making just vertex loop corrections to the bare vertices does not ameliorate the external momentum growth of scattering diagrams in the UV in our example, see Eq.~(\ref{eq:quad}). In fact, it makes the growth increase. In the next subsection, we shall dress the vertices by making both propagator and vertex loop corrections to the bare vertices at the left- and right-ends of the scattering diagrams.

\subsection{Dressing the vertices by making propagator \& vertex loop corrections to the bare vertices}\label{sec:dpv1}

\begin{figure}[t]
\centering
\includegraphics[width=.40\textwidth]{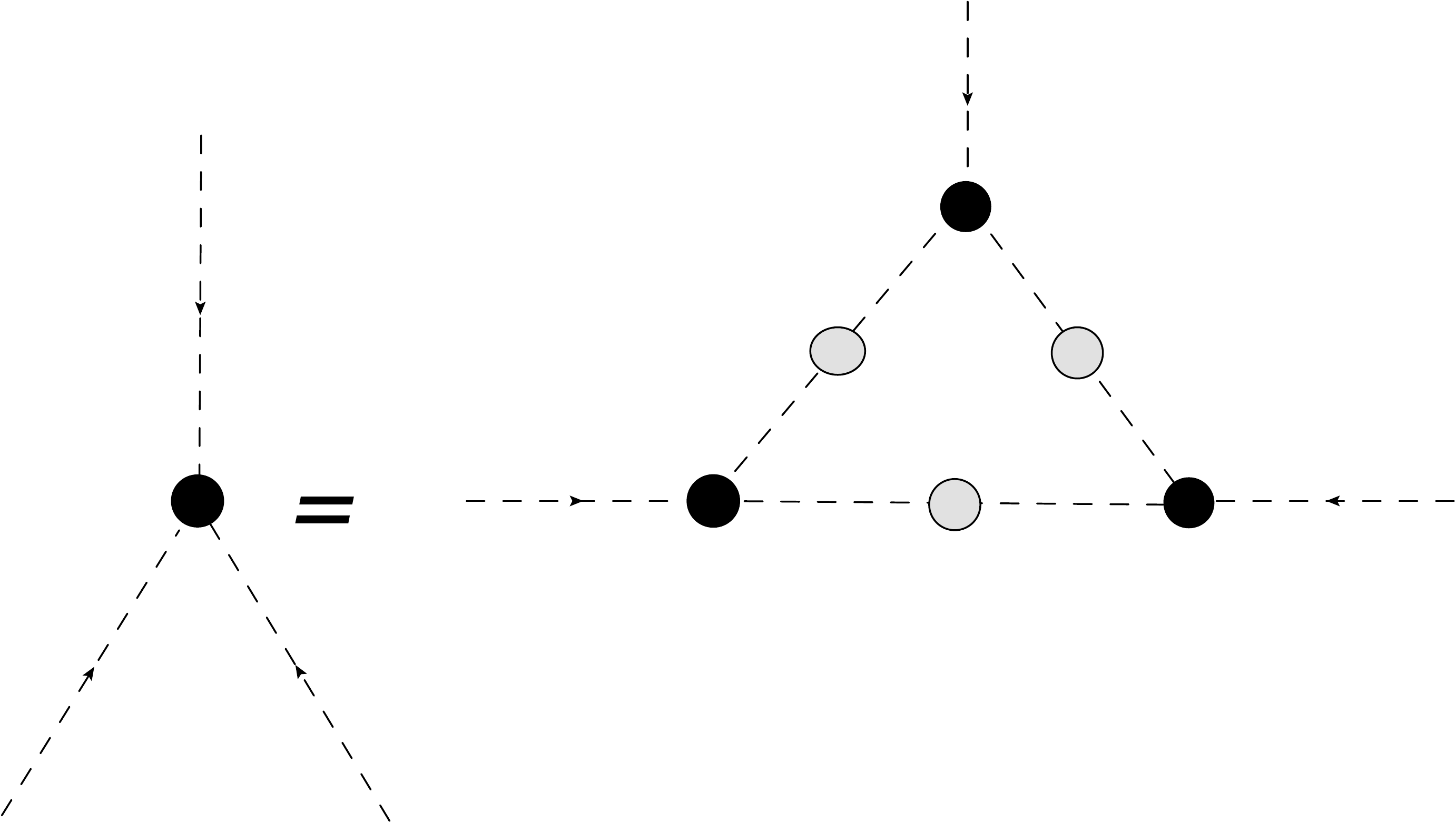}
\caption{\label{fig:set} {\small $3$-point diagram constructed out of lower-loop $2$-point \& $3$-point diagrams. The shaded blobs indicate dressed internal propagators and the dark blobs indicate renormalised vertex corrections. The loop order of the dark blob on the left is $n$ while the loop order of the dark blobs on the right is $n-1$. The external momenta are $p_1$, $p_2$, $p_3$ and the internal (that is, inside the loop) momenta are $k+\frac{p_{1}}{3}-\frac{p_2}{3}$, $k+\frac{p_{2}}{3}-\frac{p_3}{3}$, $k+\frac{p_{3}}{3}-\frac{p_1}{3}$.}}
\end{figure}

In this subsection, we shall dress the vertices by making renormalised propagator and vertex loop corrections to the bare vertices at the left- and right-ends of the scattering diagrams, see Fig.~\ref{fig:set}. Again, we expect the external momentum dependence of the $3$-point function to be given in the UV limit, {\it i.e.}, as $p_{i}\ra \infty$, where $i=1,2,3$, by Eq.~\eqref{eq:popopo}.

\begin{figure}[t]
\centering
\includegraphics[width=.40\textwidth]{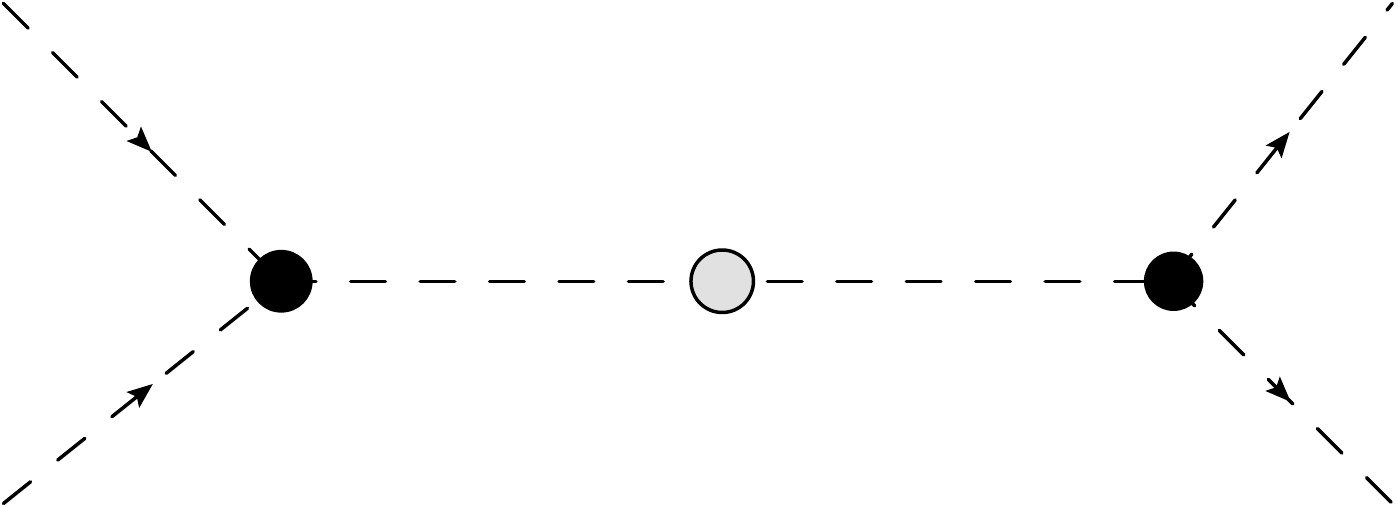}
\caption{\label{fig:achacha} {\small An $s$-channel scattering diagram $p_{1} p_{2} \ra p_{3} p_{4}$. The shaded blob indicates a dressed propagator and the dark blobs indicate renormalised vertex corrections.}}
\end{figure}

As previously, the best way to obtain the largest exponents for the external momenta is to have the $\al$ exponent correspond to the external momenta. Assuming a symmetric distribution of $(\bt, \ga)$ among the internal loops and considering the $n$-loop, $3$-point diagram with symmetrical routing of momenta,  see Fig.~\ref{fig:set}, the (dressed) propagators in the $1$-loop triangle are
\be
-i\LF k + \frac{p_{1}}{3} - \frac{p _ {2}}{3} \RF^{-8}\,, -i\LF k + \frac{p_{2}}{3} - \frac{p _ {3}}{3} \RF^{-8}\,, -i\LF k + \frac{p_{3}}{3} - \frac{p _ {1}}{3} \RF^{-8}\,, 
\ee
while the vertex factors are 
\ba
&& i p _ {1} ^ {2\al^{n-1}} \LF k + \frac{p _ {3}}{3} - \frac{p _ {1}}{3} \RF^{2\bt^{n-1}} \LF k+\frac{p _ {1}}{3} - \frac{p _ {2}}{3} \RF ^{2\ga^{n-1}}\,, \non 
&& i p _ {2} ^ {2\al^{n-1}} \LF k + \frac{p _ {1}}{3} - \frac{p _ {2}}{3} \RF^{2\bt^{n-1}} \LF k + \frac{p _ {2}}{3} - \frac{p _ {3}}{3} \RF ^{2\ga^{n-1}}\,,\non && i p _ {3} ^ {2\al^{n-1}} \LF k + \frac{p _ {2}}{3} - \frac{p _ {3}}{3} \RF^{2\bt^{n-1}} \LF k + \frac{p _ {3}}{3} - \frac{p _ {1}}{3} \RF ^{2\ga^{n-1}}\,.
\ea 
Conservation of momenta then yields, in the UV, {\it i.e.}, as $p_{i} \ra \infty$, where $i=1,2,3$,
\begin{align}\label{eq:ww}
\Ga_{3,n}&{\longrightarrow}\int \frac{\mathrm{d} ^ 4 k}{(2 \pi) ^ 4} \, \LT \frac{p_{1}^{2\al^{n-1}}p_{2}^{2\al^{n-1}}p_{3}^{2\al^{n-1}}}{\LF k + \frac{p_{1}}{3} - \frac{p _ {2}}{3} \RF^{8}\LF k + \frac{p_{2}}{3} - \frac{p _ {3}}{3} \RF^{8}\LF k + \frac{p_{3}}{3} - \frac{p _ {1}}{3} \RF^{8}} \Rd \non
&  \times \Ld \LF k + \frac{p_{1}}{3} - \frac{p _ {2}}{3} \RF^{2(\bt^{n-1}+\ga^{n-1})} \LF k + \frac{p_{2}}{3} - \frac{p _ {3}}{3} \RF^{2(\bt^{n-1}+\ga^{n-1})}\LF k + \frac{p_{3}}{3} - \frac{p _ {1}}{3} \RF^{2(\bt^{n-1}+\ga^{n-1})}\vphantom{\frac{p_{1}^{2\al^{n-1}}p_{2}^{2\al^{n-1}}p_{3}^{2\al^{n-1}}}{\LF k + \frac{p_{1}}{3} - \frac{p _ {2}}{3} \RF^{2}\LF k + \frac{p_{2}}{3} - \frac{p _ {3}}{3} \RF^{2}\LF k + \frac{p_{3}}{3} - \frac{p _ {1}}{3} \RF^{2}}}\RT\,,
\end{align}
where $p_{1}$, $p_{2}$, $p_{3}$ are the external momenta for the $1$-loop triangle and the superscript in the $\al,\bt,\ga$ indicates that these are coefficients that one obtains from contributions up to $n-1$  loop level.
Now, let us proceed to obtain the $n$-th loop coefficients by inspecting Eq.~(\ref{eq:ww}),
we have
\be
\al^n=\bt^n=\ga^n=\al^{n-1}+2(\bt^{n-1}+\ga^{n-1})\,.
\ee
For the $3$-point bare vertices, we have that $\al^{0}=\bt^{0}=1$ and $\ga^{0}=0$. As $n$ increases, $\al^n$, $\bt^n$ and $\ga^n$ increase; this means that, as the number of loops increases, the external momentum growth of the dressed vertices increases. 

If we now dress the vertices by making renormalised propagator and vertex loop corrections to the bare vertices at the left- and right-ends of the tree-level scattering diagrams, see Fig.~\ref{fig:achacha}, we obtain, as $s \ra - \infty$,
\begin{align}
\mathcal{T}_{\mathrm{both~corrections}}^{\mathrm{s-channel}}& \sim s^{2\al^{n}} \LF \frac{s}{2} \RF ^{4\al^{n}} \frac{1}{s^{4}} \,,\\
\mathcal{T}_{\mathrm{both~corrections}}^{\mathrm{t-channel}}& \sim t^{2\al^{n}} \LF \frac{s}{2} \RF ^{4\al^{n}} \frac{1}{t^{4}}= \LT \frac{s}{2}(1-\cos \theta) \RT ^{2\al^{n}-4} \LF \frac{s}{2} \RF ^{4\al^{n}} \,,\\
\mathcal{T}_{\mathrm{both~corrections}}^{\mathrm{u-channel}}& \sim u^{2\al^{n}} \LF \frac{s}{2} \RF ^{4\al^{n}} \frac{1}{u^4}=\LT \frac{s}{2}(1+\cos \theta) \RT ^{2\al^{n}-4} \LF \frac{s}{2} \RF ^{4\al^{n}} \,.
\end{align} 
Since $\al^{n} \geq 3$ for $n \geq 1$, we can make the following observations:
\begin{itemize}

\item{$\mathcal{T}_{\mathrm{both~corrections}}^{\mathrm{s-channel}}$ blows up as $s \ra - \infty$.}

\item{$\mathcal{T}_{\mathrm{both~corrections}}^{\mathrm{t-channel}}$ blows up as $s \ra -\infty$ except when $\cos(\theta)=1 \Rightarrow \theta=0$.}

\item{Similarly, $\mathcal{T}_{\mathrm{both~corrections}}^{\mathrm{u-channel}}$ blows up as $s \ra -\infty$ except when $\cos(\theta)=-1 \Rightarrow \theta=\pi$.}

\end{itemize}
Thus, one can check that the cross section $\sa_{\mathrm{both~corrections}}$ corresponding to $\mathcal{T}_{\mathrm{both~corrections}}=\mathcal{T}_{\mathrm{both~corrections}}^{\mathrm{s-channel}}+\mathcal{T}_{\mathrm{both~corrections}}^{\mathrm{t-channel}}+\mathcal{T}_{\mathrm{both~corrections}}^{\mathrm{u-channel}}$ blows up as $s \ra - \infty$.  


We see that dressing the vertices by making propagator and vertex loop corrections to the bare vertices cannot ameliorate the UV external momentum growth of scattering diagrams in our toy model example, Eq.~(\ref{eq:quad}). This motivates us to consider something very different; in the following section, we shall not consider a finite-order, higher-derivative theory, but an infinite-derivative massless scalar field theory with cubic interaction in $\phi$. Both the free and interaction parts of the action will contain infinite derivatives.

\section{Scatterings in infinite-derivative theory}\label{sec:infdev}
\numberwithin{equation}{section}

We saw in section~\ref{sec:motiv} that, within the context of a finite-order higher-derivative scalar toy model, we cannot tame the UV external momentum growth appearing in scattering diagrams. In particular, we need to ``soften'' the external momentum contributions coming from the dressed vertices; as we saw in subsections~\ref{sec:dv1}~\&~\ref{sec:dpv1}, dressing the vertices in a finite-order higher-derivative toy model cannot help us tame the external momentum growth of the scattering diagrams. Since this is not possible for a finite-order higher-derivative toy model, we shall examine an infinite-derivative scalar toy model. Therefore, let us consider the following action, which has a cubic interaction where $\lambda\ll {\cal O}(1)$, and treat it perturbatively:
\be\label{sid}
S=S_{\mt{free}}+S_{\mt{int}}\,,
\ee
where
\be
S_{\mt{free}}=\frac{1}{2}\int d^4 x \, \LF  \phi \Box a(\Box) \phi\RF
\ee
and 
\be
S_{\mt{int}}=\la \int d^4 x \, (\phi \Box \phi a(\Box) \phi)\,.
\ee
Now, let us demand that the propagator for free action retains  {\it only} the massless scalar degree of freedom. In which case, we assume that the kinetic 
term obtains an {\it entire function} correction. For simplicity, we take such a function to be Gaussian:
\be \label{eq:exp}
a (\Box) = e^{- \Box / M ^ {2}}\,,
\ee
where $M$ is the mass scale at which the non-local modifications become important. With this choice, the infinite-derivative theory can be made ghost-free~\cite{Biswas:2011ar,Biswas:2013kla}. Such a choice of  $a(\Box)$ is also well motivated by $p$-adic strings~\cite{Freund:1987kt}. 
The propagator in momentum space is then given in the Euclidean space, by
\be
\Pi (k ^ 2)= \frac{- i}{k^2 e ^ {\kb ^ 2}}\,,
\ee
where barred $4$-momentum vectors denote $\bar k = k/M$. The vertex factor for three incoming momenta $k_{1},~k_{2},~k_{3}$ satisfying the following conservation law,
\be
k _ {1} + k _ {2} + k _ {3} = 0\,,
\ee
is given by
\be \label{eq:uaua}
\la V (k _ {1}, k _ {2}, k _ {3}) = -i \la \LT k_{1}^{2}(e^{\kb_{2}^2}+e^{\kb_{3}^2})+ k_{2}^{2}(e^{\kb_{3}^2}+e^{\kb_{1}^2})+ k_{3}^{2}(e^{\kb_{1}^2}+e^{\kb_{2}^2}) \RT \,.
\ee
We can compute the tree-level $s$, $t$, $u$ channels in the CM frame and we obtain
\be
i \mathcal{T}_{\mathrm{tree-level}}^{\mathrm{s-channel}}=-\la^{2} s^{2} \LT 3e^{-s/2M^{2}}+e^{-s/M^2} \RT^{2} \LF \frac{i}{se^{-s/M^{2}}} \RF\,,
\ee
\be
i \mathcal{T}_{\mathrm{tree-level}}^{\mathrm{t-channel}}=-\la^{2} \LT (s+2t)e^{-s/2M^{2}}+se^{-t/M^2} \RT^{2} \LF \frac{i}{te^{-t/M^{2}}} \RF\,,
\ee\be
i \mathcal{T}_{\mathrm{tree-level}}^{\mathrm{u-channel}}=-\la^{2} \LT (s+2u)e^{-s/2M^{2}}+se^{-u/M^2} \RT^{2} \LF \frac{i}{ue^{-u/M^{2}}} \RF\,.
\ee
We note that, as $s \ra -\infty$, $\mathcal{T}_{\mathrm{tree-level}}=\mathcal{T}_{\mathrm{tree-level}}^{\mathrm{s-channel}}+\mathcal{T}_{\mathrm{tree-level}}^{\mathrm{t-channel}}+\mathcal{T}_{\mathrm{tree-level}}^{\mathrm{u-channel}} $ blows up. 

Now, in order to compute the dressed propagator, we have, first, to write down the $1$-loop, $2$-point function with external momenta $p$ and $-p$. We have
\begin{align}
\Ga_{2,1}(p^2) & = \frac{i \la^2}{2} \int \frac{d^4k}{(2\pi)^4} \, \frac{1}{(\frac{p}{2}+k)^{2}(\frac{p}{2}-k)^{2}e^{(\frac{\pb}{2}+\kb)^2}e^{(\frac{\pb}{2}-\kb)^2}} \non
& \times \LT \vphantom{(\frac{p}{2}-k)^{2}} p^{2}(e^{(\frac{\pb}{2}+\kb)^2}+e^{(\frac{\pb}{2}-\kb)^2})+(\frac{p}{2}+k)^{2}(e^{\pb^2}+e^{(\frac{\pb}{2}-\kb)^2})+(\frac{p}{2}-k)^{2}(e^{\pb^2}+e^{(\frac{\pb}{2}+\kb)^2})\RT^{2}\,.
\end{align}
After renormalising the divergent (in terms of the internal loop momentum $k^{\mu}$) terms~\footnote{Within the context of dimensional regularisation, we obtain an $\epsilon^{-1}$ pole, where $\epsilon=4-d$ and $d$ is the number of dimensions.}, we have that the most divergent part (in terms of the external momentum $p^{\mu}$) of the $1$-loop, $2$-point function is given by
\be
\frac{i\lambda ^2 M^4 e^{\frac{3 \pb^2}{2}} \left(4 M^2+p^2\right)}{32 \pi ^2 p^2}\,.
\ee
Thus, the renormalised $1$-loop, $2$-point function goes as $(1+4\pb^{-2})e^{\frac{3\pb^2}{2}}$ when $p$ is large.

As $s \ra -\infty$, $\Ga_{2,1\mt{r}}(-s)$ goes as $e^{-\frac{3s}{2M^2}}$. Since
\begin{align}
i\mathcal{T}_{\mathrm{1-loop}} &=  \la^{2} (p_{1},p_{2},-p_{1}-p_{2}) V(-p_{3},-p_{4},p_{1}+p_{2}) \LF \frac{i}{s e^{-s / M^2}} \RF ^ {2} \Ga_{2,1\mt{r}}(-s)  \non
&+ \la^{2} V(p_{1},-p_{3},p_{3}-p_{1}) V(p_{2},-p_{4},p_{1}-p_{3}) \LF \frac{i}{t e^{-t / M^2}} \RF ^ {2} \Ga_{2,1\mt{r}}(-t)  \non
&+ \la^{2} V(p_{1},-p_{4},p_{4}-p_{1}) V(p_{2},-p_{3},p_{1}-p_{4}) \LF \frac{i}{u e^{-u / M^2}} \RF ^ {2}\Ga_{2,1\mt{r}}(-u) \,,
\end{align}
we have that the $s$-channel of $\mathcal{T}_{\mathrm{1-loop}}$, see again Fig.~\ref{fig:kuk}, goes as $e^{-\frac{2s}{M^2}} e^{\frac{2s}{M^2}} e^{-\frac{3s}{2M^2}} = e^{-\frac{3s}{2M^2}}$ when $s \ra -\infty$. Hence, as $s \ra - \infty$, $\mathcal{T}_{\mathrm{1-loop}}^{\mathrm{s-channel}}$ diverges. $\mathcal{T}_{\mathrm{1-loop}}^{\mathrm{t-channel}}$ and $\mathcal{T}_{\mathrm{1-loop}}^{\mathrm{u-channel}}$ also diverge as $s \ra -\infty$.

\subsection{Dressing the propagator}

Since for large $p$, the dressed propagator goes as
$$\w{\Pi}(p^2) \approx (1+4\pb^{-2})^{-1} e^{-\frac{3\pb^2}{2}}\,$$
we observe that the dressed propagator is more strongly exponentially suppressed than the bare propagator.

Since
\begin{align}
i\mathcal{T}_{\mathrm{dressed}} &= \la^{2} V(p_{1},p_{2},-p_{1}-p_{2}) V(-p_{3},-p_{4},p_{1}+p_{2}) \w{\Pi}(-s) \non
&+ \la^{2}  V(p_{1},-p_{3},p_{3}-p_{1}) V(p_{2},-p_{4},p_{1}-p_{3}) \w{\Pi}(-t)  \non
&+ \la^{2}  V(p_{1},-p_{4},p_{4}-p_{1}) V(p_{2},-p_{3},p_{1}-p_{4}) \w{\Pi}(-u) \,,
\end{align}
then, if we now replace the bare propagator with the dressed propagator in the tree-level scattering diagrams, see Fig.~\ref{fig:dressed} (bottom), we will have, as $s \ra -\infty$,
\begin{align}
\mathcal{T}_{\mathrm{dressed}}^{\mathrm{s-channel}}& \sim \LT 3e^{-\frac{s}{2M^{2}}}+e^{-\frac{s}{M^2}} \RT^{2}e^{\frac{3s}{2M^2}} \sim e^{-\frac{s}{2M^2}} \,,\\
\mathcal{T}_{\mathrm{dressed}}^{\mathrm{t-channel}}& \sim \LT (s+2t)e^{-\frac{s}{2M^{2}}}+se^{-\frac{t}{M^2}} \RT^{2} e^{\frac{3t}{2M^2}}=\LT s(2-\cos \theta)e^{-\frac{s}{2M^{2}}}+se^{-\frac{s(1-\cos \theta)}{2M^2}} \RT^{2} e^{\frac{3s(1-\cos \theta)}{4M^2}}  \,,\\
\mathcal{T}_{\mathrm{dressed}}^{\mathrm{u-channel}}& \sim \LT (s+2u)e^{-\frac{s}{2M^{2}}}+se^{-\frac{u}{M^2}} \RT^{2} e^{\frac{3u}{2M^2}}=\LT s(2+\cos \theta)e^{-\frac{s}{2M^{2}}}+se^{-\frac{s(1+\cos \theta)}{2M^2}} \RT^{2} e^{\frac{3s(1+\cos \theta)}{4M^2}} \,.
\end{align}
Hence, we can make the following observations:
\begin{itemize}

\item{$\mathcal{T}_{\mathrm{dressed}}^{\mathrm{s-channel}}$ blows up as $s \ra - \infty$.}

\item{$\mathcal{T}_{\mathrm{dressed}}^{\mathrm{t-channel}}$ blows up as $s \ra -\infty$ for all values of $\theta$.}

\item{$\mathcal{T}_{\mathrm{dressed}}^{\mathrm{u-channel}}$ blows up as $s \ra -\infty$ for all values of $\theta$.}

\end{itemize}

Since $\mathcal{T}_{\mathrm{dressed}}=\mathcal{T}_{\mathrm{dressed}}^{\mathrm{s-channel}}+\mathcal{T}_{\mathrm{dressed}}^{\mathrm{t-channel}}+\mathcal{T}_{\mathrm{dressed}}^{\mathrm{u-channel}}$, one can verify that the total cross section $\sigma_{\mathrm{dressed}}$ corresponding to $\mathcal{T}_{\mathrm{dressed}}$ blows up as $s \ra - \infty$. We also observe that the external momentum dependence of $\mathcal{T}_{\mathrm{dressed}}$ exhibits less growth for large external momenta as compared to the external momentum dependence of $\mathcal{T}_{\mathrm{tree-level}}$ (or $\mathcal{T}_{\mathrm{1-loop}}$). 

To conclude, the use of the dressed propagator ameliorates the external momentum growth of the scattering diagrams, but this is not sufficient by itself. In subsection~\ref{sec:aiai}, we shall dress the vertices to see whether we can eliminate the external momentum growth of the scattering diagrams.


\subsection{Dressing the vertices by making vertex loop corrections to the bare vertices}\label{sec:aiai}

In this subsection, we shall dress the vertices by making renormalised vertex loop corrections to the bare vertices at the left- and right-ends of the scattering diagrams, see Fig.~\ref{fig:set22}. We have that both the bare propagators and the bare vertices can be written as exponentials in momenta; after integration with respect to the internal loop momentum $k$, we obtain an exponential expression where the exponents are in terms of the external momenta $p_1$, $p_2$, $p_3$. As the loop-order increases, the $3$-point function can still be written as an exponential function of the external momenta; this happens because, as previously, the (bare) propagators are exponentials in momenta while the (dressed) vertices are also exponentials in momenta.Thus, in the UV limit, {\it i.e.}, as $p_{i}\ra \infty$, where $i=1,2,3$, the $3$-point function, again see Fig.~\ref{fig:set22}, can be written as
\be \label{vvvv}
\Ga_3\st{UV}{\longrightarrow}\sum_{\al,\bt,\ga} e^{\al\pb_{1}^{2}+\bt\pb_{2}^{2}+\ga\pb_{3}^{2}}\,,
\ee
with  the convention
\be
\al\geq\bt\geq\ga\,,
\ee
where $p_1$, $p_2$, $p_3$ are the three external momenta. The reason we expect the external momentum dependence of the $3$-point function to be given by Eq.~\eqref{vvvv} is that, once all the (lower-) loop subdiagrams have been integrated out, what remains are exponential expressions in terms of the three external momenta $p_{1}$, $p_{2}$, $p_{3}$. 

The best way to obtain the largest exponents for the external momenta is to have the $\al$ exponent correspond to the external momenta. Assuming a symmetric distribution of $(\bt, \ga)$ among the internal loops and considering the $n$-loop, $3$-point diagram with symmetrical routing of momenta,  see Fig.~\ref{fig:set22}, the propagators in the $1$-loop triangle are given by
\be
e^{-\LF \kb + \frac{\pb_{1}}{3} - \frac{\pb _ {2}}{3} \RF^2}\,, e^{-\LF \kb + \frac{\pb_{2}}{3} - \frac{\pb _ {3}}{3} \RF^2}\,, e^{-\LF \kb + \frac{\pb_{3}}{3} - \frac{\pb _ {1}}{3} \RF^2}\,,
\ee
and the vertex factors are
\ba
&& e ^ {\al^{n-1} \pb _ {1} ^ {2} + \bt^{n-1} \LF \kb + \frac{\pb _ {3}}{3} - \frac{\pb _ {1}}{3} \RF^2 + \ga^{n-1} \LF \kb + \frac{\pb _ {1}}{3} - \frac{\pb _ {2}}{3} \RF ^2}\,, \non
&& e ^ {\al^{n-1} \pb _ {2} ^ {2} + \bt^{n-1} \LF \kb + \frac{\pb _ {1}}{3} - \frac{\pb _ {2}}{3} \RF^2 + \ga^{n-1} \LF \kb + \frac{\pb _ {2}}{3} - \frac{\pb _ {3}}{3} \RF ^2}\,, \non
&& e ^ {\al^{n-1} \pb _ {3} ^ {2} + \bt^{n-1} \LF \kb + \frac{\pb _ {2}}{3} - \frac{\pb _ {3}}{3} \RF^2 + \ga^{n-1} \LF \kb + \frac{\pb _ {3}}{3} - \frac{\pb _ {1}}{3} \RF ^2}\,.
\ea

Conservation of momenta then yields, in the UV, {\it i.e.}, as $p_{i} \ra \infty$, where $i=1,2,3$,
\begin{align}
\label{eq:crucial33}
\Ga_{3,n}&{\longrightarrow} \int \frac{d ^ 4 k}{(2 \pi) ^ 4} \, \LT \frac{e ^ {\al^{n-1} \pb _ {1} ^ {2} + \bt^{n-1} \LF \kb + \frac{\pb _ {3}}{3} - \frac{\pb _ {1}}{3} \RF^2 + \ga^{n-1} \LF \kb + \frac{\pb _ {1}}{3} - \frac{\pb _ {2}}{3} \RF ^2}}{e^{-\LF \kb + \frac{\pb_{1}}{3} - \frac{\pb _ {2}}{3} \RF^2}e^{-\LF \kb + \frac{\pb_{2}}{3} - \frac{\pb _ {3}}{3} \RF^2}e^{-\LF \kb + \frac{\pb_{3}}{3} - \frac{\pb _ {1}}{3} \RF^2}}\Rd \non
& \Ld \times ~~ e ^ {\al^{n-1} \pb _ {2} ^ {2} + \bt^{n-1} \LF \kb + \frac{\pb _ {1}}{3} - \frac{\pb _ {2}}{3} \RF^2 + \ga^{n-1} \LF \kb + \frac{\pb _ {2}}{3} - \frac{\pb _ {3}}{3} \RF ^2}e ^ {\al^{n-1} \pb _ {3} ^ {2} + \bt^{n-1} \LF \kb + \frac{\pb _ {2}}{3} - \frac{\pb _ {3}}{3} \RF^2 + \ga^{n-1} \LF \kb + \frac{\pb _ {3}}{3} - \frac{\pb _ {1}}{3} \RF ^2}\vphantom{\frac{e ^ {\al^{n-1} \pb _ {1} ^ {2} + \bt^{n-1} \LF \kb + \frac{\pb _ {3}}{3} - \frac{\pb _ {1}}{3} \RF^2 + \ga^{n-1} \LF \kb + \frac{\pb _ {1}}{3} - \frac{\pb _ {2}}{3} \RF ^2}}{e^{-\LF \kb + \frac{\pb_{1}}{3} - \frac{\pb _ {2}}{3} \RF^2}e^{-\LF \kb + \frac{\pb_{2}}{3} - \frac{\pb _ {3}}{3} \RF^2}e^{-\LF \kb + \frac{\pb_{3}}{3} - \frac{\pb _ {1}}{3} \RF^2}}}\RT \non
& = \int \frac{d ^ 4 k}{(2 \pi) ^ 4} \, \frac{e^{\al^{n-1}(\pb_1^2+\pb_2^2+\pb_3^2)}}{e^{[1-\bt^{n-1}-\ga^{n-1}][3\kb^2+\frac{1}{3}(\pb_1^2+\pb_2^2+\pb_3^2)]} }\,,
\end{align}
where $p_{1}$, $p_{2}$, $p_{3}$ are the external momenta for the $1$-loop triangle, and  the superscript in the $\al,\bt,\ga$ indicates that these are coefficients that one obtains from contributions up to $n-1$  loop level.

Integrating Eq.~\eqref{eq:crucial33} with respect to the loop momentum $k$ and reminding ourselves that $\al^n$, $\bt^n$ and $\ga^n$ are the coefficients of $\pb_1^2$, $\pb_2^2$ and $\pb_3^2$, respectively, appearing in the exponentials in~Eq.~(\ref{vvvv}), we have
\be 
\al^n=\bt^n=\ga^n=\al^{n-1}+\frac{1}{3}(\bt^{n-1}+\ga^{n-1})-\frac{1}{3} \,.
\label{alpha-n}
\ee
In particular, for the $1$-loop, $3$-point graph, one has to use the $3$-point bare vertices (see Eq.~\eqref{eq:uaua}): $\al^0=1$ and $\bt^0=\ga^0=0$. One then obtains
\be
\al^1=\bt^1=\ga^1=\frac{2}{3} \,,
\ee
leading to an overall symmetric vertex: $e^{\frac{2}{3}(\pb_1^2+\pb_2^2+\pb_3^2)}$ and $\al ^ {1}+\bt^{1}+\ga^{1} = 2$. We observe that, as $n$ increases, $\al^n$, $\bt^n$ and $\ga^n$ increase; this means that, as the number of loops increases, the external momentum contributions of the dressed vertices become larger and larger. 

We conclude that dressing the vertices by considering just vertex loop corrections to the bare vertices does not ameliorate the external momentum growth of scattering diagrams in the UV in our toy model example Eq.~(\ref{sid}); in fact, it makes that growth increase. In the next subsection, we shall dress the vertices by considering both propagator and vertex loop corrections to the bare vertices. 

\subsection{Dressing the vertices by making propagator \& vertex loop corrections to the bare vertices} \label{sec:juju}



As our next step, let us now consider $\mathcal{T}_{\mathrm{dressed}}$~\footnote{We could equally well consider $\mathcal{T}_{\mathrm{tree-level}}$, $\mathcal{T}_{\mathrm{1-loop}}$, etc. By making renormalised propagator \& vertex loop corrections to the bare vertices at the left- and right-ends of the scattering diagram under consideration, the external momentum growth would be eliminated at sufficiently high loop order.}. We know that $\mathcal{T}_{\mathrm{dressed}}$ diverges as $s \ra - \infty$. Let us now dress the vertices by making renormalised propagator and vertex loop corrections to the bare vertices at the left- and right-ends of the diagram. Regarding the dressed propagator, we have $\w{\Pi}(p^2)\st{UV}{\longrightarrow} e^{-{3\pb^2}/{2}}$. Therefore, following the prescription given in section~\ref{sec:aiai}, the $3$-point function can again be written as an exponential function of the external momenta; this happens because, as previously, the (dressed) propagators are exponentials in momenta while the (dressed) vertices are also exponentials in momenta. Hence, in the UV limit, {\it i.e.}, as $p_{i}\ra \infty$, where $i=1,2,3$, the $3$-point function $\Ga_{3}$, see Fig.~\ref{fig:set}, is again given by Eq.~\eqref{vvvv}. As previously, the best way to obtain the largest exponents for the external momenta is to have the $\al$ exponent correspond to the external momenta. Assuming a symmetric distribution of $(\bt, \ga)$ among the internal loops and considering the $n$-loop, $3$-point diagram with symmetrical routing of momenta,  see Fig.~\ref{fig:set}, the propagators in the $1$-loop triangle are given by
\be
e^{-\frac{3}{2}\LF \kb + \frac{\pb_{1}}{3} - \frac{\pb _ {2}}{3} \RF^2}\,, e^{-\frac{3}{2}\LF \kb + \frac{\pb_{2}}{3} - \frac{\pb _ {3}}{3} \RF^2}\,, e^{-\frac{3}{2}\LF \kb + \frac{\pb_{3}}{3} - \frac{\pb _ {1}}{3} \RF^2}\,,
\ee
and the vertex factors are
\ba
&& e ^ {\al^{n-1} \pb _ {1} ^ {2} + \bt^{n-1} \LF \kb + \frac{\pb _ {3}}{3} - \frac{\pb _ {1}}{3} \RF^2 + \ga^{n-1} \LF \kb + \frac{\pb _ {1}}{3} - \frac{\pb _ {2}}{3} \RF ^2}\,, \non
&& e ^ {\al^{n-1} \pb _ {2} ^ {2} + \bt^{n-1} \LF \kb + \frac{\pb _ {1}}{3} - \frac{\pb _ {2}}{3} \RF^2 + \ga^{n-1} \LF \kb + \frac{\pb _ {2}}{3} - \frac{\pb _ {3}}{3} \RF ^2}\,, \non
&& e ^ {\al^{n-1} \pb _ {3} ^ {2} + \bt^{n-1} \LF \kb + \frac{\pb _ {2}}{3} - \frac{\pb _ {3}}{3} \RF^2 + \ga^{n-1} \LF \kb + \frac{\pb _ {3}}{3} - \frac{\pb _ {1}}{3} \RF ^2}\,.
\ea
In the UV, {\it i.e.}, as $p_{i} \ra \infty$, where $i=1,2,3$, conservation of momenta gives
\begin{align}
\label{eq:crucial}
\Ga_{3,n}&{\longrightarrow} \int \frac{d ^ 4 k}{(2 \pi) ^ 4} \, \frac{e ^ {\al^{n-1} \pb _ {1} ^ {2} + \bt^{n-1} \LF \kb + \frac{\pb _ {3}}{3} - \frac{\pb _ {1}}{3} \RF^2 + \ga^{n-1} \LF \kb + \frac{\pb _ {1}}{3} - \frac{\pb _ {2}}{3} \RF ^2}}{e^{-\frac{3}{2} \LF \kb + \frac{\pb_{1}}{3} - \frac{\pb _ {2}}{3} \RF^2}e^{-\frac{3}{2} \LF \kb + \frac{\pb_{2}}{3} - \frac{\pb _ {3}}{3} \RF^2}e^{-\frac{3}{2} \LF \kb + \frac{\pb_{3}}{3} - \frac{\pb _ {1}}{3} \RF^2}} \non
& \times e ^ {\al^{n-1} \pb _ {2} ^ {2} + \bt^{n-1} \LF \kb + \frac{\pb _ {1}}{3} - \frac{\pb _ {2}}{3} \RF^2 + \ga^{n-1} \LF \kb + \frac{\pb _ {2}}{3} - \frac{\pb _ {3}}{3} \RF ^2}e ^ {\al^{n-1} \pb _ {3} ^ {2} + \bt^{n-1} \LF \kb + \frac{\pb _ {2}}{3} - \frac{\pb _ {3}}{3} \RF^2 + \ga^{n-1} \LF \kb + \frac{\pb _ {3}}{3} - \frac{\pb _ {1}}{3} \RF ^2} \non
& = \int \frac{d ^ 4 k}{(2 \pi) ^ 4} \, \frac{e^{\al^{n-1}(\pb_1^2+\pb_2^2+\pb_3^2)}} {e^{[\frac{3}{2}-\bt^{n-1}-\ga^{n-1}][3\kb^2+\frac{1}{3}(\pb_1^2+\pb_2^2+\pb_3^2)]} }\,,
\end{align}
where $p_{1}$, $p_{2}$, $p_{3}$ are the external momenta for the $1$-loop triangle, and  the superscript in the $\al,\bt,\ga$ indicates that these are coefficients that one obtains from contributions up to $n-1$  loop level.

After integrating Eq.~\eqref{eq:crucial} with respect to the loop momentum $k$, one obtains
\be 
\al^n=\bt^n=\ga^n=\al^{n-1}+\frac{1}{3}(\bt^{n-1}+\ga^{n-1})-\frac{1}{2} \,.
\label{alpha-nn}
\ee
For the $3$-point bare vertices, we have $\al^0=1$ and $\bt^0=\ga^0=0$. Employing Eq.~\eqref{alpha-nn}, one then obtains
\be
\al^1=\bt^1=\ga^1=\frac{1}{2} \,.
\ee
We observe that $\al ^ {1}+\bt^{1}+\ga^{1} = \frac{3}{2}$. We anticipate that the exponents become smaller as the loop order becomes larger; hence, we posit that the following inequality holds:
\be
\al^{n} +\bt^{n} +\ga^{n} \leq \frac{3}{2} \,.
\label{condition2}
\ee
Using Eq.~\eqref{alpha-nn}, we see that Eq.~\eqref{condition2} is satisfied as long as the following condition is also satisfied:
\be
\al^{n-1}+\frac{1}{3}(\bt^{n-1}+\ga^{n-1})\leq 1\,.
\label{condition3}
\ee
To recap, we have shown that if, up to loop order $n-1$, Eq.~\eqref{condition3} holds, then, at loop order $n$, Eq.~\eqref{condition2} holds too. In order to conclude the recursive argument (see~\cite{Talaganis:2014ida} for more details regarding the recursive argument), we have to show that Eq.~\eqref{condition3} holds at loop order $n$ as well. Consequently, we have
\be
\al^{n}+\frac{1}{3}(\bt^{n}+\ga^{n})=\frac{5}{3}\LT\al^{n-1}+\frac{1}{3}(\bt^{n-1}+\ga^{n-1})-\frac{1}{2}\RT\leq \frac{5}{6}<1\,.
\ee
We have verified that Eq.~\eqref{condition3} does hold at loop order $n$. As a result, the loops stay finite as the loop order increases.

Now, since $\al^1=\bt^1=\ga^1=\frac{1}{2}$, and using Eq.~\eqref{alpha-nn},we obtain that, for $n=2$,
\be
\al^2=\bt^2=\ga^2=\frac{1}{3}\,,
\ee
for $n=3$,
\be
\al^3=\bt^3=\ga^3=\frac{1}{18}\,,
\ee
for $n=4$,
\be
\al^4=\bt^4=\ga^4=-\frac{11}{27}\,.
\ee

We conclude that, for $n \geq 4$, $\al^n$, $\bt^n$ and $\ga^n$ become negative. The fact that $\al^n$, $\bt^n$ and $\ga^n$ become negative for sufficiently large $n$ should be emphasised since it is precisely this negativity which eliminates the external momentum growth of the scattering diagrams in the UV.

For $n=4$, we have the following results:
\begin{itemize}

\item{We find that the largest external momentum contribution of the $s$-channel, see Fig.~\ref{fig:achacha}, goes as
\be
e^{\frac{44s}{27M^{2}}}e^{\frac{3s}{2M^{2}}}=e^{\frac{169s}{54M^{2}}}\,,
\ee
which tends to $0$ as $s \ra -\infty$.} 

\item{Regarding the $t$-channel, the largest external momentum contribution goes as
\be
e^{\frac{22t}{27M^2}}e^{\frac{22s}{27M^2}}e^{\frac{3t}{2M^2}}=e^{\frac{s(213-125\cos \theta)}{108M^2}}\,,
\ee
which, again, tends to $0$ as $s \ra -\infty$ for all values of $\theta$.}

\item{Regarding the $u$-channel, the largest external momentum contribution goes as
\be
e^{\frac{22u}{27M^2}}e^{\frac{22s}{27M^2}}e^{\frac{3u}{2M^2}}=e^{\frac{s(213+125\cos \theta)}{108M^2}}\,,
\ee
which, again, tends to $0$ as $s \ra -\infty$ for all values of $\theta$. Hence, for sufficiently large $n$ (specifically, for $n \geq 4$), there is no exponential growth for the $s$-, $t$- and $u$-channels as $s \ra - \infty$.}

\end{itemize}

Let us also point out that we do not have to worry about polynomial growth in $s$ since any polynomial functions of $s$ will be multiplied by exponential functions of $s$ and their product will tend to $0$ as $s \ra - \infty$, keeping in mind that exponential functions always dominate polynomial ones at large values.

Dressing the vertices by making {\it both} propagator and vertex loop corrections to the bare vertices ameliorates and, in fact, completely eliminates, for sufficiently large $n$, the external momentum growth of the scattering diagrams in the UV. In the next section, we will study an infinite-derivative scalar toy model inspired by a ghost-free and singularity-free theory of gravity.


\section{Scattering in infinite-derivative theories of gravity}\label{sec:infdevgrav}
\numberwithin{equation}{section}

Inspired by the results of previous section, let us now investigate scattering diagrams in the context of infinite-derivative theories of gravity, which is ghost-free and singularity-free, for brevity we call it BGKM gravity~\cite{Biswas:2011ar}. In~\cite{Talaganis:2014ida}, we studied the quantum loops for an infinite-derivative scalar field theory action as a toy model to mimic the UV properties of the BGKM gravity. Expanding the BGKM action around the Minkowski vacuum~\footnote{One could expand the BGKM action and, subsequently, derive the propagator for a different background metric such as (A)dS~\cite{Biswas:2016etb}. Computing graviton-graviton scattering diagrams in (A)dS spacetime is a topic for future investigation.}, one can obtain, for instance,  the ``free'' part that determines the propagator from the $\cO (h^2)$ terms
; $h_{\mu \nu}$ denotes a small perturbation around Minkowski spacetime: $g_{\mu \nu}=\eta_{\mu \nu}+h_{\mu \nu}$.
The $\cO (h^3)$ terms determine cubic interaction vertices. Unfortunately, $\cO (h^3)$  terms  are technically challenging and some of the expressions involve double sums. Instead of getting involved with too many technicalities, we shall, therefore, choose to work with a simple toy model
action that respects a combination of shift and scaling symmetry at the level of equation of motion. This will  allow us to capture some of  the essential features  of BGKM gravity, such as the compensating nature of exponential suppression in the propagator and an exponential enhancement in the vertex factor.

The infinite-derivative action that can modify the propagator of the graviton  without introducing any new states  is of the form~\cite{Biswas:2011ar}
\be
\label{action}
S = S_{EH} + S_{Q}\,,
\ee
where $S_{EH}$ is the Einstein-Hilbert action,
\be
\label{eq:EH}
\int d ^ 4 x \, \sqrt{-g} \, \frac{R}{2}\,,
\ee
and $S_{Q}$ is given by
\be
\label{nlaction}
S_{Q} = \int d ^ 4 x \, \sqrt{-g} \LT R \cF _ {1} (\Box) R + R _ {\mu \nu} \cF _ {2} (\Box) R ^ {\mu \nu} + R _ {\mu \nu \lambda \sigma} \cF _ {3} (\Box) R ^ {\mu \nu \lambda \sigma}\RT\,,
\ee
where the $\cF_{i}$'s are analytic functions of $\Box$ (the covariant d'Alembertian operator):
\be
\cF_{i} (\Box) = \sum _ {n=0}^{\infty} f_{i_n} \Box^{n}\,,
\label{quadratic}
\ee
satisfying
\be
\label{eq:kuku}
2\cF_1+\cF_2+2\cF_3=0\,,
\ee
and the constraint that the combination
\be
\label{eq:a}
a (\Box) = 1 - \frac{1}{2} \mathcal{F} _ {2} (\Box) \Box - 2 \mathcal{F} _ {3} (\Box) \Box\,,
\ee
is an {\it entire function} with no zeroes. In Eq.~\eqref{quadratic}, the $f_{i_n}$'s  are real coefficients. Eqs.~\eqref{action}-\eqref{eq:a} define the BGKM gravity models. 
For BGKM gravity, we have the propagator~\cite{Biswas:2011ar,Biswas:2013kla},
\be
\Pi(k^2) = - \frac{i}{k^2a(-k^2)}\LF {\cal P}^2 - \frac{1}{2} {\cal P}_s ^0  \RF=\frac{1}{a(-k^2)} \Pi_{GR}\,,
\ee
for the physical degrees of freedom for a graviton propagating in $4$ dimensions; see~\cite{Biswas:2011ar,Biswas:2013kla,peter} for the definitions of
the spin projector operators ${\cal P}^2$ and ${\cal P}_s ^0$.

Since we know that the field equations of GR exhibit a global scaling symmetry,
\be
g_{\mu\nu}\ra \la g_{\mu\nu}\,.
\label{scaling}
\ee
When we expand the metric around the
Minkowski vacuum, 
\be
g_{\mu \nu} = \eta_{\mu \nu} + h_{\mu \nu}\,,
\label{minkowski}
\ee
the scaling symmetry translates to a symmetry for $h_{\mu\nu}$, whose infinitesimal version is given by
\be
h_{\mu\nu} \ra (1 + \epsilon) h_{\mu\nu} + \epsilon\eta_{\mu\nu}\,.
\ee
The symmetry relates the free and interaction terms just like gauge symmetry does. Thus, we are going to use this combination of shift and scaling symmetry,
\be
\phi \ra (1 + \epsilon) \phi + \epsilon\,,
\label{scale-shift}
\ee
to arrive at a scalar {\it toy model}, whose propagator and vertices preserve the compensating nature found in the full BGKM gravity.
Now, let us write down explicitly the scalar \textit{toy model} action and the Feynman rules for that action, {\it i.e.}, the propagator and the vertex factors. Our scalar {\it toy model} action is given by:
\be
\label{eq:action}
S_{\mt{scalar}} = S _ {\mt{free}} + S _ {\mt{int}}\,,
\ee
where
\be
\label{free}
S_{\mt{free}} = \frac{1}{2}\int d^4 x \, \LF  \phi \Box a(\Box) \phi\RF
\ee
and
\be
\label{int}
S_{\mt{int}} = \frac{1}{M_P} \int d ^ 4 x \, \LF \frac{1}{4} \phi \partial _ {\mu} \phi \partial ^ {\mu} \phi + \frac{1}{4} \phi \Box \phi a(\Box) \phi - \frac{1}{4} \phi \partial _ {\mu} \phi a(\Box)  \partial ^ {\mu} \phi \RF\,.
\ee
For the purpose of this paper, we are going to choose:
\be
a (\Box) = e^{- \Box / M ^ {2}}\,,
\ee
where $M$ is the mass scale at which the non-local modifications become important.
The propagator in momentum space for Eq.~(\ref{free}) is then given by
\be
\Pi (k ^ 2)= \frac{- i}{k^2 e ^ {\kb ^ 2}}\,,
\ee
where barred $4$-momentum vectors from now on will denote the  momentum divided by the mass scale $M$. The vertex factor for three incoming momenta $k_{1},~k_{2},~k_{3}$ satisfying the conservation law:
\be
\label{conservation}
k _ {1} + k _ {2} + k _ {3} = 0\,,
\ee
is then given by
\be
\label{eq:V}
\frac{1}{M_{P}}V (k _ {1}, k _ {2}, k _ {3}) = \frac{i}{M_P} C(k_1,k_2,k_3) \LT 1 -  e ^ {\kb _ {1} ^ {2}} -  e ^ {\kb _ {2} ^ {2}} - e ^ {\kb _ {3} ^ {2}}\RT\,,
\ee
where
\be
C (k_1,k_2,k_3)= \frac{1}{4} \LF k _ {1} ^ {2} + k _ {2} ^ {2} + k _ {3} ^ {2} \RF\,.
\ee
For the above set-up,  $1$-loop, $2$-point diagram, both with zero and arbitrary external momenta have been computed in Ref.~\cite{Talaganis:2014ida}, which gives a $\La^4$ divergence, where $\La$ is a momentum cut-off. Further, $1$-loop, $N$-point diagrams with vanishing external momenta were also computed. The $2$-loop diagrams with zero external momenta also give a $\La^4$ divergence, suggesting that we do not get new divergences as we proceed from $1$-loop to $2$-loop. In Ref.~\cite{Talaganis:2014ida}, the authors have computed $1$-loop and $2$-loop computations with external momenta and paid extra care in understanding the $1$-loop, $2$-point function which appeared as a subdivergence in higher-loop diagrams.

Typically, in the $1$-loop, $2$-point function, the authors obtained $e^{\frac{3 \pb^2}{2}}$ external momentum dependence in the UV, which indicates that, for $\pb^{2} \ra \infty$, the $1$-loop, $2$-point function tends to infinity. This may appear as an initial setback, but, actually, this external momentum dependence is what, we believe, makes all higher-loop and higher-point diagrams finite once the bare propagators were replaced by dressed propagators. The dressed propagator is given by (see Ref.~\cite{Talaganis:2014ida})
\be
\w{\Pi}(p^2)= \frac{\Pi(p^2)}{1-\Pi(p^2)\Ga_{2,1\mt{r}}(p^2)}= \frac{-i}{p ^ {2 } e ^ { \bar{p} ^ {2}}-\frac{M^{4}}{M_{P}^{2}}f\LF\bar{p}^2\RF}\,,
\ee
where $f(\pb^2)$ grows as $e^{\frac{3\pb^2}{2}}$ as $\pb^2 \ra \infty$. For such an external momentum dependence, the dressed propagator is more strongly suppressed than the bare one. The finiteness of all higher-loop and higher-point diagrams became possible because  the exponential suppression in the dressed propagator, which is $e^{-\frac{3\pb^{2}}{2}}$ in the UV, overcame the exponential enhancement arising from the vertices. The $1$-loop, $N$-point functions with zero external momenta became UV-finite, and so did the $2$-loop integrals for vanishing external momenta. The basic reason is simple; even for the $1$-loop diagrams, the suppression coming from the propagators is stronger than the enhancements coming from the vertices. This ensures two things - first, it makes the loops finite and, second, the UV growth of the finite diagrams with respect to the external momenta becomes weaker in every subsequent loops. Thus, finiteness of higher loops is ensured recursively.

With this adequate information, we now concentrate on the scattering problem for BGKM gravity.
We can compute the $s,~t,~u$-channels, tree-level scattering diagram $p_{1} p_{2} \ra p_{3} p_{4}$, see Fig.~\ref{fig:tree}, which is given by in the Euclidean space, as:
\be
i \mathcal{T}_{\mathrm{tree-level}}^{\mathrm{s-channel}}=\frac{1}{M_{P}^{2}} V(p_{1},p_{2},-p_{1}-p_{2}) V(-p_{3},-p_{4},p_{1}+p_{2}) \LF \frac{i}{s e^{-s / M^2}} \RF\,.
\ee
\be
i \mathcal{T}_{\mathrm{tree-level}}^{\mathrm{t-channel}}=\frac{1}{M_{P}^{2}} V(p_{1},-p_{3},p_{3}-p_{1}) V(p_{2},-p_{4},p_{4}-p_{2}) \LF \frac{i}{t e^{-t / M^2}} \RF\,,
\ee
\be
i \mathcal{T}_{\mathrm{tree-level}}^{\mathrm{u-channel}}=\frac{1}{M_{P}^{2}} V(p_{1},-p_{4},p_{4}-p_{1}) V(p_{2},-p_{3},p_{3}-p_{2}) \LF \frac{i}{u e^{-u / M^2}} \RF\,.
\ee
Therefore, we have 
\begin{align}
\mathcal{T}_{\mathrm{tree-level}} & = \frac{1}{16 M_{P}^{2} \LF p_{1}+p_{2} \RF ^{2} e^{(\pb_{1}+\pb_{2})^2}} \LT p_{1}^{2}+p_{2}^{2}+\LF p_{1}+p_{2} \RF^{2} \RT \LT p_{3}^{2}+p_{4}^{2}+\LF p_{1}+p_{2} \RF^{2} \RT \non
& \times \LT 1 - e^{\pb_{1}^{2}} - e^{\pb _{2}^{2}} - e^{(\pb_{1}+\pb_{2})^{2}}\RT \LT 1 - e^{\pb_{3}^{2}} - e^{\pb _{4}^{2}} - e^{(\pb_{1}+\pb_{2})^{2}}\RT \non
& + (p_{2}\leftrightarrow -p_{3}) \non
& + (p_{2}\leftrightarrow -p_{4})\,.
\end{align}
In the CM frame, we obtain:
\begin{align}
\mathcal{T}_{\mathrm{tree-level}} & = - \frac{1}{16 M_{P}^{2} s e^{-\frac{s}{M^2}}} \LF - 2s \RF^{2} \LF 1 - 2 e^{-\frac{s}{2M^2}} - e^{-\frac{s}{M^2}} \RF^{2} \non
& - \frac{1}{16 M_{P}^{2} t e^{-\frac{t}{M^2}}} \LF - s - t \RF^{2} \LF 1 - 2 e^{-\frac{s}{2M^2}} - e^{-\frac{t}{M^2}} \RF^{2} \non
& - \frac{1}{16 M_{P}^{2} u e^{-\frac{u}{M^2}}} \LF - s - u \RF^{2} \LF 1 - 2 e^{-\frac{s}{2M^2}} - e^{-\frac{u}{M^2}} \RF^{2}\,.
\end{align}
Let us again point out that $s$, $t$, $u$ are all negative in Euclidean space and satisfy $s=u+t$. Clearly, the cross section $\sa_{\mathrm{tree-level}}$ corresponding to $\mathcal{T}_{\mathrm{tree-level}}$ blows up as $s \ra - \infty$ since $\lvert \mathcal{T} \rvert^2$ diverges in that limit.

Before we compute the scattering amplitude, let us first consider the $1$-loop, $2$-point function, see Fig.~\ref{fig:okopoko}, with arbitrary external momenta, which is given by
\ba
\label{2pt-ext}
\Ga_{2,1}(p^2) = \frac{i}{2 i ^ {2} M _ {P} ^ {2}} \int \frac{d ^ 4 k}{(2 \pi) ^ {4}} \, \frac{V ^ {2} (-p, \frac{p}{2} + k, \frac{p}{2} - k)}{(\frac{p}{2} + k) ^ {2} (\frac{p}{2} - k) ^ {2} e ^ {\LF\frac{\pb}{2} + \kb\RF ^ {2}} e ^ {\LF\frac{\pb}{2} - \kb\RF ^ {2}}}\,.
\ea 
Using the dimensional regularisation scheme, we obtain an $\epsilon^{-1}$ pole, 
\be
\Ga_{2,1,\mt{div}}(p^2)=\frac{i p ^ {4}}{64 \pi ^ {2} M _ {P} ^ {2}}\frac{1}{\epsilon}\,,
\ee
as expected, which can be eliminated using a suitable counter-term. The counter-term, which is needed to cancel the $\epsilon^{-1}$ divergence and which should be added to the action in Eq.~\eqref{eq:action}, is given by
\be
S_{\mt{ct}}= - \frac{1}{128\en \pi ^2M_P^2}\int d^4x \, \phi \Box^2 \phi\,,
\ee
yielding
\be
\Ga_{2,1,\mt{ct}}(p^2)=-\frac{i p ^ {4}}{64 \pi ^ {2} M _ {P} ^ {2}} \frac{1}{\epsilon}\,.
\ee 
Had we employed a hard cut-off $\La$, the maximum divergence would have been $\La^4$.

Therefore, regarding the renormalised $1$-loop, $2$-point function, $\Ga_{2,1\mt{r}}$, with external momenta $p,-p$, we have $\Ga_{2,1\mt{r}} = \frac{i M^4}{M_{P}^{2}}f(\pb^2)$, where
 \begin{align}\label{dugu}
f (\pb^2) & = \frac{\pb ^ {4}}{128 \pi ^ {2}} \left(- \log \left(\frac{\pb^2}{4 \pi}\right) - \gamma +2 \right) \non
& + \frac{e^{-\pb^2}}{512 \pi ^2 \pb^2} \LT \vphantom{\text{Ei}\LF \frac{\pb^2}{2}\RF}-2 e^{\pb ^ 2} \LF e ^ {2 \pb ^2} - 1 \RF \pb^6 Ei \LF-\pb^2 \RF + \LF e ^ {\pb^2} - 1\RF \LF \vphantom{\text{Ei}\LF \frac{\pb^2}{2}\RF}-2 \LF \pb^4 + 3 \pb^2 + 2 \RF \Rd \Rd \non
& + \Ld \Ld  \LF e^{\frac{3 \pb^2}{2}} -e^{\frac{\pb^2}{2}} \RF \LF 2\pb^4 + 5 \pb^2 + 4 \RF + e ^{\pb^2} \LF e^{\pb^2} -1 \RF \pb^6 Ei \LF - \frac{\pb^2}{2} \RF + 2 e^{\pb^2} \LF 7 \LF \pb^4 + \pb^2 \RF +2 \RF \RF \RT \,.
\end{align}
Now, regarding the $1$-loop scattering diagram, see Fig.~\ref{fig:kuk}, we obtain:
\begin{align}
\mathcal{T}_{\mathrm{1-loop}} &=  V(p_{1},p_{2},-p_{1}-p_{2}) V(-p_{3},-p_{4},p_{1}+p_{2}) \LF \frac{i}{s e^{-s / M^2}} \RF ^ {2} \frac{M^4}{M_{P}^{4}} f(-s) \non
&+  V(p_{1},-p_{3},p_{3}-p_{1}) V(p_{2},-p_{4},p_{1}-p_{3}) \LF \frac{i}{t e^{-t / M^2}} \RF ^ {2} \frac{M^4}{M_{P}^{4}} f(-t) \non
&+  V(p_{1},-p_{4},p_{4}-p_{1}) V(p_{2},-p_{3},p_{1}-p_{4}) \LF \frac{i}{u e^{-u / M^2}} \RF ^ {2} \frac{M^4}{M_{P}^{4}} f(-u)\,,
\end{align}
where $\Ga_{2,1\mt{r}} = \frac{i M^4}{M_{P}^{2}}f(-s)=\frac{iM^4}{M_{P}^2}f(\pb^2)$, where $f(\pb^2)$ is given by Eq.~\eqref{dugu} and $f(\pb^2)$ is a regular analytic function of $\pb^2$ which grows as $e^{\frac{3\pb^2}{2}}$ as $\pb^2 \ra \infty$. 

As $s \ra -\infty$, $\Ga_{2,1\mt{r}}(-s)$ (and $f(-s)$) goes as $e^{-\frac{3s}{2M^2}}$. The $s$-channel of $\mathcal{T}_{\mathrm{1-loop}}$ goes as $e^{-\frac{2s}{M^2}} e^{\frac{2s}{M^2}} e^{-\frac{3s}{2M^2}} = e^{-\frac{3s}{2M^2}}$ when $s \ra -\infty$. As $s \ra - \infty$, $\mathcal{T}_{\mathrm{1-loop}}^{\mathrm{s-channel}}$ diverges. $\mathcal{T}_{\mathrm{1-loop}}^{\mathrm{t-channel}}$ and $\mathcal{T}_{\mathrm{1-loop}}^{\mathrm{u-channel}}$ also diverge.

\subsection{Dressing the propagator and the vertices}\label{sec:laolao}

Similar to the earlier cases, we have found that dressed propagator is more strongly exponentially suppressed than the bare propagator.
Since $\Pi(p^2)\Ga_{2,1\mt{r}}(p^2)$ grows with large momenta, we have, for large $p$,
\be
\label{dressed-UV}
\w{\Pi}(p^2)\ra \Ga^{-1}_{2,1\mt{r}}(p^2)\approx \LF 9-12 \pb^{-2}  \RF ^ {-1} e^{-\frac{3\bar{p}^2}{2}}\,.
\ee
Now, if we replace the bare propagator with the dressed propagator in the tree-level scattering diagrams, see Fig.~\ref{fig:dressed} (bottom), we obtain:
\begin{align}
\mathcal{T}_{\mathrm{dressed}} &= V(p_{1},p_{2},-p_{1}-p_{2}) V(-p_{3},-p_{4},p_{1}+p_{2}) \LF \frac{1}{M_{P}^{2} s e^{-s / M^2} + M^{4}f(-s)} \RF \non
&+  V(p_{1},-p_{3},p_{3}-p_{1}) V(p_{2},-p_{4},p_{1}-p_{3}) \LF \frac{1}{M_{P}^{2}t e^{-t / M^2}+ M^{4}f(-t)} \RF  \non
&+  V(p_{1},-p_{4},p_{4}-p_{1}) V(p_{2},-p_{3},p_{1}-p_{4}) \LF \frac{1}{M_{P}^{2}u e^{-u / M^2}+ M^{4}f(-u)} \RF \,,
\end{align}
where, as $s \ra -\infty$, $f(-s)$ goes as $e^{-\frac{3s}{2M^2}}$.  
An explicit computation, see Fig.~\ref{fig:dressed} (bottom), gives us, as $s \ra -\infty$,
\begin{align}
\mathcal{T}_{\mathrm{dressed}}^{\mathrm{s-channel}}& \sim \LT 2e^{-\frac{s}{2M^2}}+e^{-\frac{s}{M^2}} - 1 \RT^{2}e^{\frac{3s}{2M^2}} \sim e^{-\frac{s}{2M^2}} \,,\\
\mathcal{T}_{\mathrm{dressed}}^{\mathrm{t-channel}}& \sim \LT 2e^{-\frac{s}{2M^2}}+e^{-\frac{t}{M^2}} - 1 \RT^{2} e^{\frac{3t}{2M^2}}=\LT 2e^{-\frac{s}{2M^2}}+e^{-\frac{s(1-\cos \theta)}{2M^2}} - 1 \RT^{2} e^{\frac{3s(1-\cos \theta)}{4M^2}}  \,,\\
\mathcal{T}_{\mathrm{dressed}}^{\mathrm{u-channel}}& \sim \LT 2e^{-\frac{s}{2M^2}}+e^{-\frac{u}{M^2}}- 1 \RT^{2} e^{\frac{3u}{2M^2}}=\LT 2e^{-\frac{s}{2M^2}}+e^{-\frac{s(1+\cos \theta)}{2M^2}}- 1  \RT^{2} e^{\frac{3s(1+\cos \theta)}{4M^2}} \,.
\end{align}
Hence, we can make the following observations:
\begin{itemize}

\item{$\mathcal{T}_{\mathrm{dressed}}^{\mathrm{s-channel}}$ blows up as $s \ra - \infty$.}

\item{$\mathcal{T}_{\mathrm{dressed}}^{\mathrm{t-channel}}$ blows up as $s \ra -\infty$ for all values of $\theta$.}

\item{$\mathcal{T}_{\mathrm{dressed}}^{\mathrm{u-channel}}$ blows up as $s \ra -\infty$ for all values of $\theta$.}

\end{itemize}

Since $\mathcal{T}_{\mathrm{dressed}}=\mathcal{T}_{\mathrm{dressed}}^{\mathrm{s-channel}}+\mathcal{T}_{\mathrm{dressed}}^{\mathrm{t-channel}}+\mathcal{T}_{\mathrm{dressed}}^{\mathrm{u-channel}}$, one can verify that the total cross section $\sigma_{\mathrm{dressed}}$ corresponding to $\mathcal{T}_{\mathrm{dressed}}$ blows up as $s \ra - \infty$. We also observe that the external momentum dependence of $\mathcal{T}_{\mathrm{dressed}}$ grows less for large external momenta as compared to the external momentum dependence of $\mathcal{T}_{\mathrm{tree-level}}$ (or $\mathcal{T}_{\mathrm{1-loop}}$). Hence, the use of the dressed propagator ameliorates the external momentum growth of the scattering diagrams, but it is not sufficient by itself.
 
To see whether we can eliminate the external momentum growth of the scattering diagrams, we will dress the vertices by making renormalised vertex loop corrections to the bare vertices at the left- and right-ends of the scattering diagrams, see Fig.~\ref{fig:set22}. Following exactly the same prescription as in section~\ref{sec:aiai}, we obtain the relation
\be 
\al^n=\bt^n=\ga^n=\al^{n-1}+\frac{1}{3}(\bt^{n-1}+\ga^{n-1})-\frac{1}{3} \,,
\ee
which is Eq.~\eqref{alpha-n}. Since $\al^0=1$ and $\bt^0=\ga^0=0$, we observe that the coefficients $\al^n$, $\bt^n$ and $\ga^n$ increase as $n$ increases; thus, dressing the vertices by keeping the propagators bare and making just vertex loop corrections to the bare vertices at the left- and right-ends of the scattering diagrams cannot tame the external momentum growth of the scattering diagrams. 


For that reason, and as an example, we will now dress the bare vertices at the left- and right-ends of the scattering diagram whose scattering matrix element is $\mathcal{T}_{\mathrm{dressed}}$ by making {\it both} propagator and vertex loop corrections to the said vertices, see Fig.~\ref{fig:set}. Following the same reasoning as in section~\ref{sec:juju}, $\al^n$, $\bt^n$ and $\ga^n$ become negative for $n \geq 4$. 

For $n=4$, we have the following conclusions:
\begin{itemize}

\item{As in section~\ref{sec:juju}, the largest external momentum contribution of the $s$-channel, see Fig.~\ref{fig:achacha}, goes as
\be
e^{\frac{44s}{27M^{2}}}e^{\frac{3s}{2M^{2}}}=e^{\frac{169s}{54M^{2}}}\,,
\ee
which tends to $0$ as $s \ra -\infty$.} 

\item{The largest external momentum contribution of the $t$-channel goes as
\be
e^{\frac{22t}{27M^2}}e^{\frac{22s}{27M^2}}e^{\frac{3t}{2M^2}}=e^{\frac{s(213-125\cos \theta)}{108M^2}}\,,
\ee
which, again, tends to $0$ as $s \ra -\infty$ for all values of $\theta$.}

\item{The largest external momentum contribution of the $u$-channel goes as
\be
e^{\frac{22u}{27M^2}}e^{\frac{22s}{27M^2}}e^{\frac{3u}{2M^2}}=e^{\frac{s(213+125\cos \theta)}{108M^2}}\,,
\ee
which, again, tends to $0$ as $s \ra -\infty$ for all values of $\theta$. Hence, for sufficiently large $n$ (specifically, for $n \geq 4$), there is no exponential growth for the $s$-, $t$- and $u$-channels as $s \ra - \infty$. The external momentum growth of $\mathcal{T}_{\mathrm{tree-level}}$, $\mathcal{T}_{\mathrm{1-loop}}$ etc. would also be eliminated following this prescription at sufficiently high loop order.}

\end{itemize}

We observe that, for sufficiently large $n$, dressing the vertices by making both propagator and vertex loop corrections to the bare vertices at the left- and right-ends of the scattering diagrams makes the external momentum dependence of any scattering diagram convergent in the UV. By considering renormalised propagator and vertex loop corrections to the bare vertices, we can eliminate the external momentum growth appearing in scattering diagrams in the regime of large external momenta, {\it i.e.}, as $s \ra -\infty$. In contrast, dressing the vertices by considering just vertex loop corrections to the bare vertices is not sufficient. Thus, dressing the vertices by making both propagator and vertex loop corrections to the bare vertices is essential to taming the external momentum growth of scattering diagrams in the UV and, as a result, we expect the cross sections of those diagrams to be finite (see Eq.~\eqref{eq:a4} in appendix~\ref{sec:defconv} for the relation between the differential cross section $d \sa$ and the scattering matrix element $\mathcal{T}$).

\section{Conclusions}\label{sec:concl}
\numberwithin{equation}{section}

The aim of this paper has been to examine the external momentum dependence of scattering diagrams in the context of infinite-derivative field theories and gravity. We have found that for a finite-order, higher-derivative scalar field theory the cross section of tree-level scattering diagrams  blows up at large momenta. Even considering 
dressed propagators and dressed vertices, by making propagator and vertex loop corrections to the bare vertices of the scattering diagrams, is not sufficient to eliminate the 
external momentum growth. However, we have noticed that dressing the propagators indeed ameliorates the external momentum growth a bit. Motivated by these results, we studied an infinite-derivative, non-local scalar field theory with non-local interactions. In this setup, the propagators are exponentially suppressed and the vertices are exponentially enhanced. 

For such non-local interactions, we have found that the tree-level cross section still blows up in the UV. Also, dressing the propagator is not sufficient to tame the 
growth. On the other hand, dressing the bare vertices by making renormalised propagator and vertex loop corrections to the bare vertices at sufficiently high loop order (when the loop order $n$ satisfies $n\geq 4$) can potentially yield finiteness of the cross section in the UV. What leads to this conclusion is the {\it softening} of the vertices. At higher loop order, the dressed vertices lead to negative exponents, which effectively softens any high-energy scattering amplitude.  As a result, the scattering cross section is expected not to blow up for large external momenta, which is encouraging as to the infinite-derivative theories of gravity under consideration.
We may speculate that, for such cases, scattering scalar wave packets with non-local interactions would not lead to black hole singularity. 
This is indeed an interesting result which can help us to understand the UV properties of gravity, if gravity itself were treated non-locally in the UV.

This motivates us to study high-energy scattering diagrams in a scalar toy-model inspired by the non-local, singularity-free theory of gravity introduced in Ref.~\cite{Talaganis:2014ida}. In this
case, we were able to demonstrate that dressing the vertices and the propagators indeed leads to a cross section that is expected to be finite for the scattering diagrams, which become convergent in the ultraviolet. This gives rise to a very interesting possibility that perhaps our recipe could be followed for pure gravity, as in the case of BGKM, to show that such non-locality indeed softens 
the trans-Planckian scattering problem and can avoid forming a black hole singularity. We believe that  our results will have consequences for understanding problems such as black hole singularity and the cosmological singularity problem in a time-dependent setup.

\section{Acknowledgements}

The authors would like to thank Tirthabir Biswas, Aindri\'u Conroy, Valery Frolov, Steven Giddings, Alex Koshelev, Warren Siegel,  Ali Teimouri, Terry Tomboulis for helpful discussions. ST is supported by a scholarship from the Onassis Foundation and AM is supported by the STFC grant ST/J000418/1.

\setcounter{equation}{0}
\section{Appendix}
\appendix

\section{Definitions and Conventions in Euclidean Space}\label{sec:defconv}
\numberwithin{equation}{section}

Let us define $s= - \LF p_{1}+p_{2} \RF ^ {2} = - \LF p_{3}+p_{4} \RF ^ 2=-E_{\mathrm{CM}}^2$, where $p_{1}+p_{2}=p_{3}+p_{4}$. Moreover, $t = - \LF p_{1}-p_{3} \RF ^ {2}= - \LF p_{2}-p_{4} \RF ^ 2$ and $u = - \LF p_{1}-p_{4} \RF ^{2} = - \LF p_{2}-p_{3} \RF ^{2}$. We have that $s$, $t$, $u$ are all negative in Euclidean space and satisfy $s=u+t$. We should keep in mind that we consider massless particles in this paper and, in Minkowski space (``mostly plus'' metric signature), $p_{i}^{2}=-m_{i}^{2}=0$, where $i=1,2,3,4$.

The total cross section, $\sigma$, in the centre-of-mass (CM) frame is given by
\be \label{eq:sigma}
\sigma = \frac{1}{S} \int _ {t_{\mathrm{min}}} ^ {t_{\mathrm{max}}} d t \,  \frac{d \sigma}{d t}\,,
\ee
where $t_{\mathrm{min}}$ and $t_{\mathrm{max}}$ are given by
\be
t= -2 E_{1}E_{3}+2\lvert \mathbf{p}_{1} \rvert \lvert \mathbf{p}_{3} \rvert \cos \theta\,,
\ee
with $\cos \theta =-1$ and $+1$, respectively ($\theta$ is the angle between $\lvert \mathbf{p}_{1} \rvert$ and $\lvert \mathbf{p}_{3} \rvert$). $S$ is the symmetry factor for $n_{i}^{'}$ identical outgoing particles of type $i$,
\be
S=\prod_{i}n_{i}^{'}! \,,
\ee
and, for two outgoing particles (after we analytically continue to Euclidean space), we have
\be \label{eq:a4}
\frac{d \sigma}{d t}= - \frac{1}{64 \pi s \lvert \mathbf{p}_{1} \rvert ^2} \lvert \mathcal{T} \rvert^{2}\,,
\ee
where $\mathcal{T}$ is the scattering matrix element. In the CM frame, we also have
\be
\lvert \mathbf{p}_{1} \rvert= \lvert \mathbf{p}_{2} \rvert =\lvert \mathbf{p}_{3} \rvert =\lvert \mathbf{p}_{4} \rvert = E_{1} = E_{2} =E_{3} =E_{4} =\frac{\sqrt{-s}}{2}\,.
\ee

Furthermore, we have that $t_{\mathrm{min}}=s$ and $t_{\mathrm{max}}=0$. Since the two outgoing particles are identical, the symmetry factor is $S=2$. Moreover, in Euclidean space,
\be
\label{t}
t= \frac{s}{2} \LF 1 - \cos \theta \RF
\ee
and
\be
\label{u}
u = \frac{s}{2} \LF 1 + \cos \theta \RF\,.
\ee

\end{document}